\documentclass[12pt]{article}
\pdfoutput=1
\usepackage{graphicx,amsmath,amsfonts,amssymb,stmaryrd,latexsym,url,subcaption,hyperref,mathtools,float,cite}

\begin{document}

\baselineskip=18pt \pagestyle{plain} \setcounter{page}{1}

\begin{center}

{\Large \bf  Benefits of marriage as a search strategy} \\ [9mm]

{\normalsize \bf Davi B. Costa \\ [3mm]
{\small \it Physics Department, University of Chicago, Chicago, IL 60637, USA}}


\end{center}

\vspace*{0.2cm}

\begin{abstract}
We propose and investigate a model for mate searching and marriage in large societies based on a stochastic matching process and simple decision rules. Agents have preferences among themselves given by some probability distribution. They randomly search for better mates, forming new couples and breaking apart in the process. Marriage is implemented in the model by adding the decision of stopping searching for a better mate when the affinity between a couple is higher than a certain fixed amount. We show that the average utility in the system with marriage can be higher than in the system without it. Part of our results can be summarized in what sounds like a piece of advice: don't marry the first person you like and don't search for the love of your life, but get married if you like your partner more than a sigma above average. We also find that the average utility attained in our stochastic model is smaller than the one associated with a stable matching achieved using the Gale-Shapley algorithm. This can be taken as a formal argument in favor of a central planner (perhaps an app) with the information to coordinate the marriage market in order to set a stable matching. To roughly test the adequacy of our model to describe existent societies, we compare the evolution of the fraction of married couples in our model with real-world data and obtain good agreement. In the last section, we formulate the model in the limit of an infinite number of agents and find an analytical expression for the evolution of the system.
\end{abstract}

\newpage
\tableofcontents

\newpage
\section{Introduction}

Consider the following situation: you have a mate and wonder whether you should propose marriage or not. There are many aspects of marriage to consider, but you are mainly concerned with marriage as a search strategy. You like your mate, but you are unsure if it is enough. After all, there may be someone out there that you would like even more, but it is unlikely that you will get to know this person if you get married. Perhaps, the best is to stay open to the possibility of meeting a potential better mate by not getting married. However, if your relationship is symmetric in this respect, you may become single because your mate left you for someone else. So while marriage closes off some of the opportunities that you would otherwise have, on the other hand, it protects you from the risk of being left out. So you ask yourself: as a search strategy, is it beneficial to get married? Or does the potential benefit of meeting a person with whom you have a greater affinity outweigh the chances of being alone because your current partner has left you?

A famous problem related to this subject is the stable matching problem \cite{10.2307/2312726}. It is the problem of finding a stable matching between two groups of agents with preferences among themselves. If we interpret the groups as man and woman, then a matching is a bijection between these two genders, and a matching is stable if there are no man and woman who both prefer each other over their respective mate. Since its inception, there has been extensive research on the mathematical structure of stable matching and related algorithmic questions \cite{10.2307/2312726,knuthstable,gusfield1989stable,roth1992two}. The notion of stable matching is fundamental to understand a variety of markets, as one can find in the recent review \cite{ren2021matching}. Furthermore, matching theory is a relevant tool for market design \cite{article}. However, when it comes to large societies, the concept of stable matching may not be important. After all, it is likely that there are two agents that would rather be together than be with their respective mates. But, in a large society, this blocking pair may not yet know each other. Also, the likelihood that this will happen depends on the type of relationship they have. If they are married, in principle, they will not be open to the possibility of matching someone else. Otherwise, they may eventually match a better mate. To make these ideas precise and to answer the question from the previous paragraph, we propose a model for dating and marriage based on a random matching process and simple decision rules.

The model contains a number $N$ of agents with preferences among themselves sampled by a Gaussian distribution. Initially, all agents are singles, and the system evolves in a certain number $T$ of discrete steps. At each step, they will match in pairs by random chance. If the agents of the matched pair like each other more than their previous mates, or more than being single, they will form a new couple. An agent becomes single if its partner starts a new relationship. Marriage is implemented in the model by adding the decision of stopping searching for a better mate when an agent finds another one with an affinity higher than a certain fixed value $\Lambda$. This parameter should be interpreted as exogenous in the sense that it characterizes the decision behavior of the agents of a given society with respect to marriage. We simulate this system for different values of $\Lambda$ and we computed the average utility in each case as a function of the number of steps. Part of the conclusions one can extract from the results we obtained, answers the question from the first paragraph. The answer sounds like a piece of advice: don't marry the first person you like and don't search for the love of your life, but get married if you like your partner more than a sigma above average.



The idea of using formal tools for the study of marriage is not new \cite{10.2307/1831130,10.2307/1829987,10.2307/1837421}. Following the recent taxonomy for models of the marriage market proposed in \cite{doi:10.1146/annurev-economics-012320-121610}, our model fits best in the category of search models (see \cite{10.2307/23045893,10.1257/002205105775362014,doi:10.1146/annurev-economics-111809-125046,10.1257/jel.20150777} for a detailed presentation of the work in search theory, and  \cite{10.2307/2780247,10.2307/2951279,Burdett1999LongTermPF,10.2307/2999430,10.2307/44955185,RePEc:bpj:bejmac:v:advances.1:y:2001:i:1:n:5} for examples of search models of marriage). However, our approach is very different from the more traditional one in economics. We investigate the step-by-step evolution of our system of agents by randomly matching them and applying the simple decision rules mentioned in the previous paragraph. We are particularly interested in the limit $N\gg T$, i.e., when the number of agents is much larger than the number of times agents match one another. We start from an initial state with everyone single to an advanced stage with a fraction of agents in a married or unmarried relationship. Then, search frictions come from the interplay between the benefit of finding a better mate and the cost of being left. As done in \cite{RePEc:bpj:bejmac:v:advances.1:y:2001:i:1:n:5}, we did not concentrate on the steady-state as in most investigations in search theory. Our non-standard approach suggests a rough empirical test of our model: we can compare the fraction of married couples at each step in our model with the fraction of persons of a given generation that are married at a certain age. We did this test by comparing Fig. (\ref{mcsimulation}) and Fig. (\ref{mcrealdata}), which arguably gives a good agreement when taken into account the purposeful simplicity of our model.

Another famous problem somewhat related to the one we are interested in here is the secretary problem (see \cite{10.2307/2245639} for a review). Very briefly, the secretary problem is an optimal stopping problem, where applicants for a position are presented sequentially, and it is necessary to choose the best candidate constrained by the condition that an applicant cannot be chosen after being rejected. In its simplest form, the parameter that characterizes applicants is sampled by the same distribution. This feature, however, is one of the essential differences between the secretary problem and the problem we are considering. The marriage market has externalitie, which are manifested in the fact that the probability distributions that samples potential mates changes over time. In principle, with these distributions, which we will compute in section (\ref{analytical}), one could formulate our model as a generalization of the secretary problem.

The paper is organized as follow: in section (\ref{dynamicalmatching}) we will present our model and in (\ref{benefitsofmarriage}) the results obtained in our simulations. In section (\ref{discussionandcomments}), we will answer some possible objections to the results and we will compare the evolution of married couples in our model and real-world data. In section (\ref{analytical}) we will formulate our model in the large $N$ limit giving an analytical formulation and generalizing many assumptions.
The steps of evolution for the analytical setup will be worked out in sections (\ref{firststep}) and (\ref{nextstep}), and in (\ref{marriedcouplesdistribution}) we will derive the asymptotic $t\rightarrow\infty$ limit for the model with marriage. 
A Mathematica notebook with the material in this article can be accessed in \cite{notebook}.


\section{Model and results}
\label{model}

As mentioned in the introduction, we are particularly interested in exploring the limit $N\gg T$. That is, when the number of agents that makes society is much larger than the number of times agents match with one another. We will see that in this limit the important degrees of freedom are not individual preferences but their probability distribution among society. 

\subsection{Stochastic matching model for mate searching}
\label{dynamicalmatching}

The model contains $N$ agents with preferences among themselves described by a utility function \cite{mas1995microeconomic}. This information is encoded in the off-diagonal elements of the \textit{affinity matrix} $A\in\mathbb{R}^{N\times N}$, where $A_{ij}$ is the utility for agent $i$ to be in a couple with agent $j$. Note that $A$ is not symmetric in general because the utility for agent $i$ to be with agent $j$ is not necessarily the same as the reverse. In addition, $A_{ij}$ is not positive in general, as it's common to find people you'd rather be single than date. 
We will often refer to the collection of agents of our model as the society.

\begin{figure}
\begin{subfigure}{.48\textwidth}
  \centering
  \includegraphics[width=\linewidth]{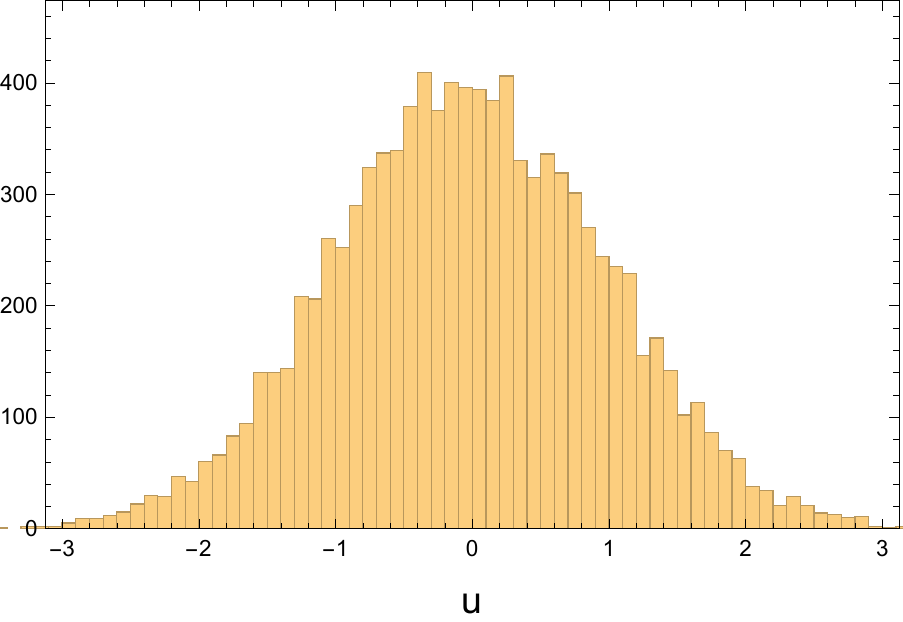}
  \caption{Gaussian model $\mathcal{N}(0,1)$.}
  \label{InitialGaussianHistogram}
\end{subfigure}%
\hspace{5mm}
\begin{subfigure}{.48\textwidth}
  \centering
  \includegraphics[width=\linewidth]{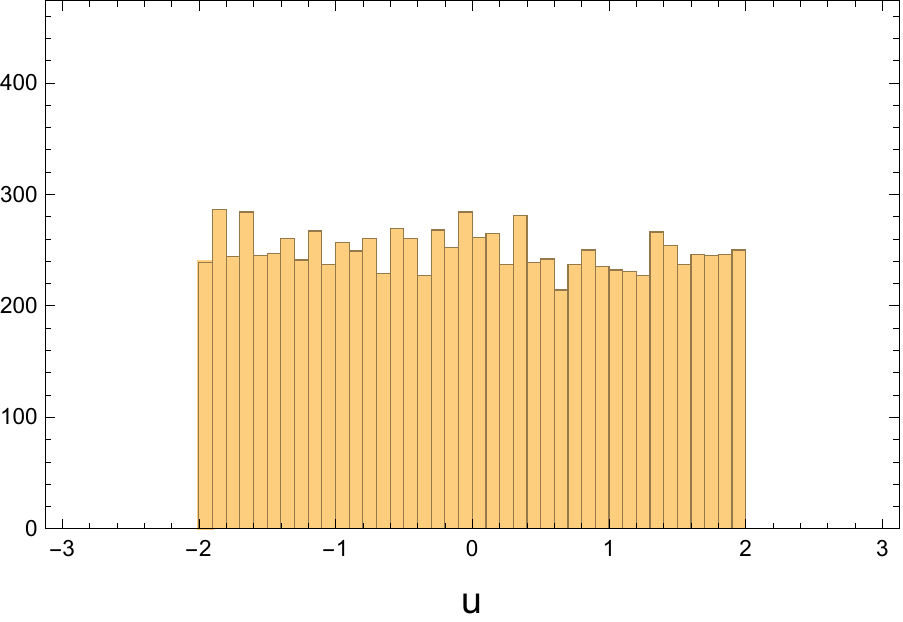}
  \caption{Uniform model $U(-2,2)$.}
  \label{InitialUniformHistogram}
\end{subfigure}
\caption{Histogram of the values $\{A_{kj}:1\leq j\leq N\textrm{ and }j\neq k\}$ of the affinity matrix for some agent $k$ with bin size $0.1$ in the Gaussian (\ref{InitialGaussianHistogram}) and Uniform (\ref{InitialUniformHistogram}) model for $N=10.000$, $\sigma=1$ and $\mu=0$.}
\label{InitialHistograms}
\end{figure}

In large societies, the probability distribution that generates the affinity matrix is more important than its actual values. Indeed, the preferences of agent $k$ concerning the other agents of society are encoded by $\{A_{kj}:1\leq j\leq N\textrm{ and } j\neq k\}$. The histogram with these values, as in Fig. (\ref{InitialHistograms}), will distribute according to some probability distribution. As it happens with many human traits, it may be reasonable to assume that these values are generated by a Gaussian distribution, i.e., $A_{kj}\sim\mathcal{N}(\mu_k,\sigma_k)$ for all $1\leq j\leq N$ with $j\neq k$ and each $1\leq k\leq N$. In principle, one may think that it is more realistic to assume that the preferences of different agents are given by distributions defined with different parameters $\mu$ and $\sigma$. Note, however, that the values of the $k-$column of the affinity matrix, $\{A_{jk}:1\leq j\leq N\textrm{ and } j\neq k\}$, encodes the way agent $k$ is liked by society. Therefore, if agents indeed have preferences given by different Gaussian distributions, then the distribution that generates the columns of the affinity matrix will not in general be Gaussian. Hence, to have both lines and columns consistently Gaussian, we may assume that all off-diagonal entries of the affinity matrix are generated by the same Gaussian distribution, that is $A_{kj}\sim\mathcal{N}(\mu,\sigma)$ for all $1\leq j\leq N$ and $1\leq k\leq N$ with $j\neq k$. When this is the case, we will say that we are dealing with the Gaussian model. For comparison and to make concrete our analytical investigations of section (\ref{analytical}), we will also study the Uniform model, i.e., $A_{ij}\sim U(\mu-2\sigma,\mu+2\sigma)$.

The diagonal elements of the affinity matrix are the utilities when agents are singles. We also believe it is reasonable to assume these values are generated by a Gaussian distribution. 
Note, that the mean of the diagonal elements gives the average utility when all agents are single. On the other hand, the mean of the off-diagonal elements can be interpreted as the average utility when agents form random couples. Therefore, the relationship between the mean of these two distributions encodes how being single compares to dating a random agent in society. We do not believe this two quantities are generically the same. But, for concreteness, we will present our results assuming they are. However, we will show in section (\ref{discussionandcomments}) that our conclusions still hold when this is not the case. In addition, in our analytical formulation of section (\ref{analytical}), these distributions will be generic.

The dynamics of the system happen in discrete steps $t\in\mathbb{N}$. At each step $t$, all agents match at random with a new agent. More precisely, at each step, there is a random matching between all agents. That is, a bijection

\begin{align}
    \mu_t:\{1,2,\dots,N\} & \rightarrow \{1,2,\dots,N\}
    \label{matching}
\end{align}

\noindent generated by a random process with uniform distribution. The matching $\mu_t$ maps $k$ to $\mu_t(k)$, the agent $k$ match at $t$. Suppose first, that agent $k$ is single at $t$ having $u_t^k=A_{kk}$, and match with agent $l=\mu_{t+1}(k)$ at $t+1$. These two agents will form a couple if they like themselves, more precisely we will have

\begin{align}
    u_{t+1}^k=\begin{cases} A_{kl}, & A_{kl}>u_t^k=A_{kk}\ \textrm{and }\ A_{lk}>u_{t}^l,\\
    u_t^k, & \textrm{otherwise}.
    \end{cases}
    \label{singleevolution}
\end{align}

\noindent Now, suppose agent $k$ is in a couple with agent $m$ at $t$. In this case $u_t^k=A_{km}$ and $u_t^m=A_{mk}$. At step $t+1$, let agent $k$ match with agent $l=\mu_{t+1}(k)$ and $m$ with agent $n=\mu_{t+1}(m)$. Then, if the affinity among matched agents is larger then the affinity between the mates, a new couple will be formed and the old couples will break apart. More precisely

\begin{align}
    u_{t+1}^k=\begin{cases} A_{kl}, & \textrm{if}\quad A_{kl}>u_t^k\ \textrm{and }\ A_{lk}>u_{t}^l,\\
    A_{kk}, & \textrm{if}\quad A_{mn}>u_t^m\ \textrm{and }\ A_{nm}>u_t^n,\\
    u_t^k, & \textrm{otherwise}.
    \end{cases}
    \label{coupleevolution}
\end{align}

\noindent We start the system with all agents as singles, that is $u_0^k=A_{kk}$ for all $1\leq k\leq N$. We should emphasize that here, a couple is not a married couple. We will introduce marriage in the next section.

Running the simulation by $T$ steps produces a $N\times T$ matrix $\{u_t^k:1\leq k\leq N,1\leq t\leq T\}$ that gives the utility of agent $k$ at each step $t$. As we are interested in large societies $T\ll N$, we are more interested in the average utility and its variance:

\begin{align}
    u_t\equiv\frac{1}{N}\sum_{k=1}^Nu_t^k,\qquad \sigma_t\equiv\sqrt{\frac{1}{N}\sum_{k=1}^N(u_t^k)^2-\Big(\frac{1}{N}\sum_{k=1}^Nu_t^k\Big)^2}
\end{align}

\noindent (in fact, all moments carry aggregate information of the system, but we will postpone to section (\ref{analytical}) the treatment of higher moments). The marked difference between $u_t^k$ and $u_t$ can be visualized by plotting their values for a simulation of the Gaussian $\mathcal{N}(0,1)$ model with $N=10.000$:

\begin{figure}[H]
    \centering
    \begin{minipage}{0.49\textwidth}
        \centering
        \includegraphics[width=\textwidth]{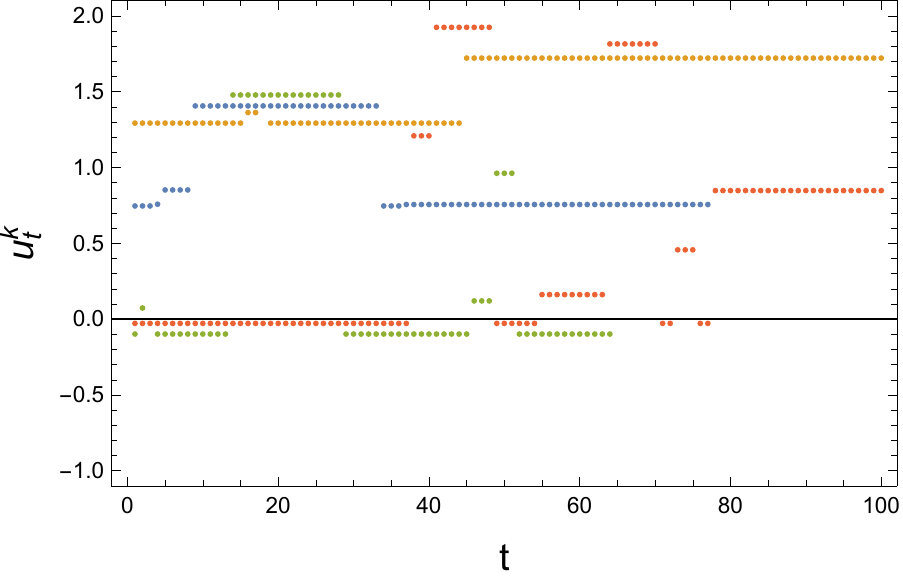} 
        \caption{Evolution of the utility of four agents, distinguished by color, in the Gaussian model $\mathcal{N}(0,1)$ with $N=10.000$}
        \label{uktGn10000s1l1}
    \end{minipage}\hfill
    \begin{minipage}{0.49\textwidth}
        \centering
        \includegraphics[width=\textwidth]{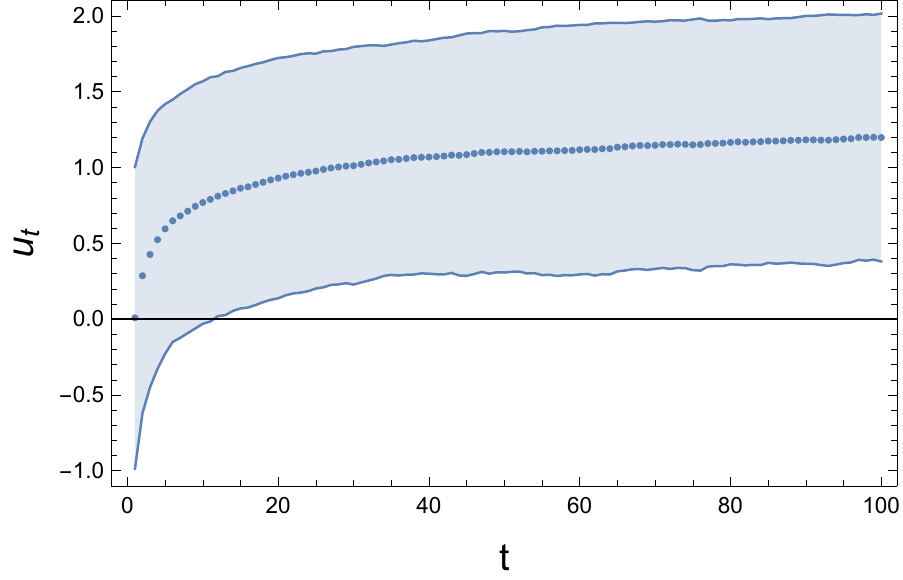}
        \caption{Evolution of average utility with band equal to the variance in the Gaussian model $\mathcal{N}(0,1)$ with $N=10.000$}
        \label{utGn10000s1l1}
    \end{minipage}
\end{figure}

\noindent The plot for the Uniform model $U(-2,2)$, which has variance $\sigma=1$, is very similar and hardly distinguishable visually from the above ones.

As in Fig. (\ref{uktGn10000s1l1}) and Fig. (\ref{utGn10000s1l1}), the results we will present were obtained in simulations using $\mu=0$ and $\sigma=1$. However, for the case that all entries of the affinity matrix are given by the same distribution, there is a symmetry that allow us to translate the results to generic $\mu$ and $\sigma$. Note, that the dynamical equations (\ref{singleevolution}) and (\ref{coupleevolution}), depends on inequalities between values of the affinity matrix such as $A_{ij}>A_{kl}$. This fact implies that the dynamics of the system will be the same if we shift and re-scale all values of the affinity matrix $A_{ij}\rightarrow\tilde{A}_{ij}=\sigma A_{ij}+\mu$. But this corresponds to using generic $\mu$ and $\sigma$ instead of 0 and 1. By doing so, the utility of each agent at each step will shift and re-scale accordingly, that is $u_t^k\rightarrow \sigma u_t^k+\mu$. Hence we can translate the results obtained for $\mu=0$ and $\sigma=1$ to generic $\mu$ and $\sigma$. In symbols, we have:

\begin{align}
    \textrm{if}\qquad(\mu=0,\sigma=1)\mapsto u_t^k\qquad\textrm{then}\qquad (\mu,\sigma)\mapsto \sigma u_t^k+\mu.
    \label{generalmusigma}
\end{align}

\noindent Using this dictionary we can easily translate the results of Fig. (\ref{uktGn10000s1l1}) and Fig. (\ref{utGn10000s1l1}) to general $\mu$ and $\sigma$. We will have $u_t^k\rightarrow \sigma u_t^k+\mu$ and $u_t\rightarrow \sigma u_t+\mu$. Therefore, it suffices to think of $0$ as $\mu$ and interpret the units of the vertical axis as given in terms of $\sigma$.

As stressed before, in large societies the number of matchings we have during our lifetimes is much smaller than the number of agents that make our societies $T\ll N$. In particular, the asymptotic limit $T\rightarrow\infty$ doesn't seem to be important. This resonates with the famous Keynes quote ``in the long run we are all dead" \cite{keynes1923tract}, which can be read as the advice that asymptotic limits and equilibrium states are not always relevant. In addition, the asymptotic limit $T\rightarrow\infty$ does not approach any equilibrium state in this model. Imagine for simplicity 3 agents, $A,B$ and $C$. Suppose $A$ has higher affinity with agent $C$, $B$ with $C$, and $C$ with $A$. At each step one of the agents does not match and they will remain in an eternal dance, which is not uncommon in modern relationships. The fact that this system never stabilizes shows that even generalized concepts of stability \cite{Multi-period,doval2021dynamically,Credibility,DAMIANO200534} may not be important to understand the matching of couples in large societies.

\subsection{Adding the decision of getting married}
\label{benefitsofmarriage}

We can implement the decision of getting married in our stochastic matching model in the following way. If the affinity between a couple is higher than a certain fixed value $\Lambda\in\mathbb{R}$, then the couple will not match with new agents in the next steps of the evolution. In this case, we say that the couple is married. More precisely, 

\begin{align}
    \textrm{the $(k,l)$-couple is married at $t$ if}\qquad u^k_t=A_{kl}>\Lambda\qquad \textrm{and}\qquad u^l_t=A_{lk}>\Lambda.
    \label{marriedcoupled}
\end{align}

\noindent Let $\mathcal{M}_{t-1}\subseteq\{1,2,\dots N\}$ be the set of agents married at time $t-1$, then the matching at time $t$ will not include them

\begin{align}
    \mu_t^\Lambda:\{1,2,\dots,N\}-\mathcal{M}_{t-1}\rightarrow\{1,2,\dots,N\}-\mathcal{M}_{t-1},
    \label{marriedcouples}
\end{align}

\noindent where $\{1,2,\dots,N\}-\mathcal{M}_{t-1}=\mathcal{M}^c_t$ is the complement of $\mathcal{M}_t$ over $\{1,2,\dots,N\}$, that is, the set of unmarried agents. Note that for each value of $\Lambda$ we will have a different simulation. Therefore, in this model, marriage is a one real parameter family. We should interpret $\Lambda$ as exogenous in the sense that it characterizes the decision behavior of the agents of a given society. Different societies will have different values of $\Lambda$. This will become clear with the discussion following the comparison between Fig. (\ref{mcsimulation}) and Fig. (\ref{mcrealdata}) in the next section. A natural generalization of this setup is attained by having a different value of $\Lambda$ for each agent, i.e., $\{\Lambda_k:1\leq k\leq N\}$. Assuming that $\Lambda_k\sim\mathcal{N}(\Lambda,\sigma_\Lambda)$ for all $1\leq k\leq N$, as we did for the elements of the affintiy matrix, then the case we are investigating can be thought of as the $\sigma_\Lambda\rightarrow0$ limit. 

As in Eq. (\ref{generalmusigma}), results obtained using $\mu=0$, $\sigma=1$ can be translated to results obtained for generic $\mu$ and $\sigma$. However, from Eq. (\ref{marriedcoupled}), one sees that it is necessary to shift and re-scale $\Lambda$ to preserve the dynamic. More precisely, let $u_t^k$ be the utility of agent $k$ obtained for $\mu=0$, $\sigma=1$ and $\Lambda$. Then, the simulation for generic $\mu$ and $\sigma$, will give $\sigma u_t^k+\mu$ if one uses $\sigma\Lambda+\mu$ instead of $\Lambda$. In symbols we have:

\begin{align}
    \textrm{if}\qquad (\mu=0,\sigma=1,\Lambda)\mapsto u_t^k\qquad\textrm{then}\qquad (\mu,\sigma,\sigma\Lambda+\mu)\mapsto \sigma u_t^k+\mu.
    \label{dictionarymarried}
\end{align}

\noindent We see that our results are still generic for $\mu$ and $\sigma$ with the appropriate transformation of $\Lambda$.

\begin{figure}
\centering\includegraphics{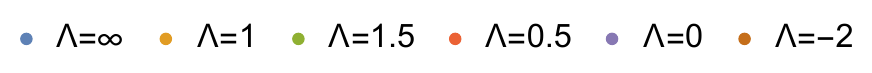}
\begin{subfigure}{.49\textwidth}
  \centering
  \includegraphics[width=\linewidth]{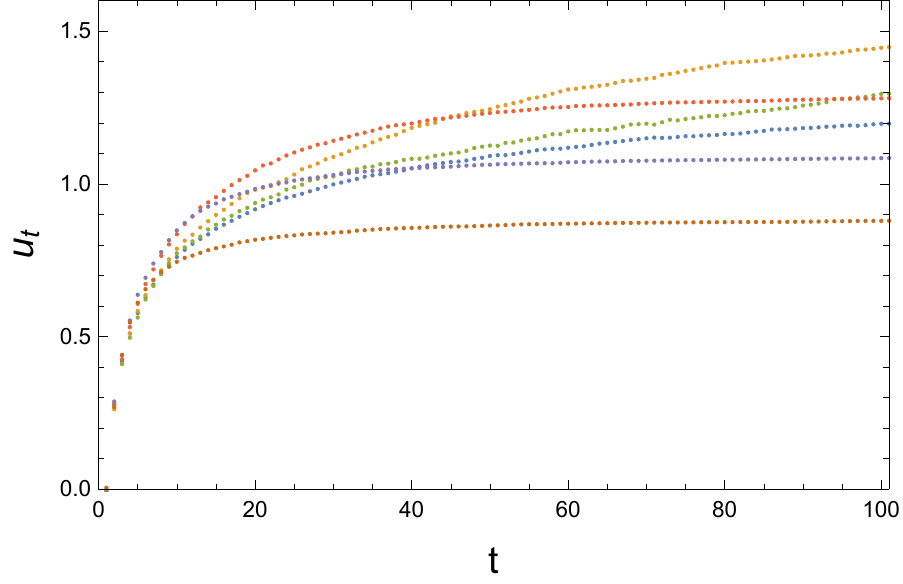}
  \caption{Gaussian model $\mathcal{N}(0,1)$.}
  \label{lcompUG10000s1}
\end{subfigure}%
\begin{subfigure}{.49\textwidth}
  \centering
  \includegraphics[width=\linewidth]{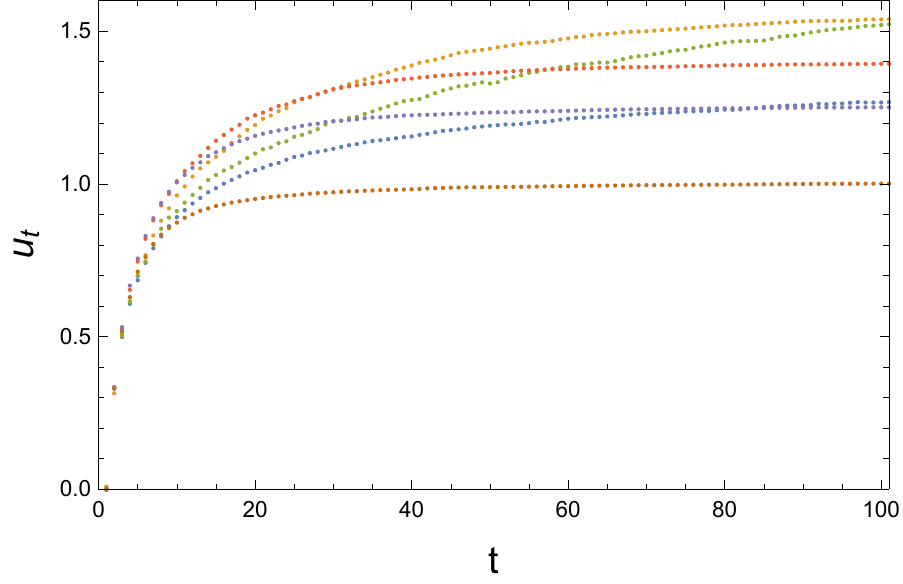}
  \caption{Uniform model $U(-2,2)$.}
  \label{lcompUN10000s1}
\end{subfigure}
\caption{Average utility evolution in the Gaussian and Uniform model with $N=10.000$ for different values of $\Lambda$. We used $\Lambda=\infty$ to denote the system without marriage.}
\label{lcomp}
\end{figure}

We can compare the average utility in the system with and without marriage for different values of $\Lambda$. Our results are presented in Fig. (\ref{lcompUG10000s1}) and Fig. (\ref{lcompUN10000s1}). The most significant lessons from our stochastic model can be extracted from these figures. Note that if we have $\Lambda>A_{ij}$ for all $i,j$, then marriage will not affect the dynamics of the system. Therefore, for sufficiently large $\Lambda$ (and finite $N$), the system with marriage evolves the same way as the system without it. On the other hand, if $\Lambda<A_{ij}$ for all $i,j$, then every agent will marry with the first matched agent they like, and the average utility will get smaller as one can see from the plots. The lesson we can extract from this result is that we should not get married to the first person we like. However, for some intermediary values of $\Lambda\simeq\sigma$, the average utility is consistently larger in the system with marriage. Therefore, it is reasonable to conclude that we should marry if we like our partner more than a sigma above average. For $\Lambda\simeq\sigma$, the lines cross, therefore, there will be a different value $\Lambda_t$ that maximizes the average utility at different steps $t$. 

To help further clarify the possible interpretations of the results, consider the following scenario. During our lives, we can take all opportunities we have to match others, pursuing the dream of finding the greatest love of our lives. Alternatively, after finding someone who makes us happy more than a certain amount, we can decide to stop this search. From the results of this model we may conclude that, on average, it is better to stick with someone good enough than always be looking for the big love of our lives. The conjunction of these results justifies the sentence we advertised before: don't marry the first person you like and don't search for the love of your life, but get married if you like your partner more than a sigma above average.

There is a second interpretation for these results that personally motivated this model. Suppose you are dating, and you should decide if you will engage in monogamy or non-monogamy. The most salient difference is that in non-monogamy, you and your mate would be open to match with others. To decide, you need to consider several features of your preferences. For example, how much you like to match new people periodically. Whether you have enough time to spend with your mate and sporadic matches. How jealous you feel when your mate match with others. To mention but a few. In addition, the openness characteristic of non-monogamous relations increases the chances of matching with someone either you or your mate have more affinity. Hence, you may wonder if the odds of finding someone better pays the cons of being left out by your mate. If one interprets what we called a couple as non-monogamy, and what we called a married couple as monogamy, then the model proposed addresses this question and gives a negative answer. The significant conclusion is that, if you are in all respects unsure whether you prefer monogamy or non-monogamy, the benefit of marriage as a search strategy disclosed by this model should tip the balance in favor of monogamy. Perhaps, in the future, this interpretation will be more significant as the number of people that experience non-monogamous relationships is growing \cite{doi:10.1080/0092623X.2016.1178675}. However, throughout the paper, we will stick to the first interpretation.

\begin{figure}
\begin{subfigure}{.49\textwidth}
  \centering
  \includegraphics[width=\linewidth]{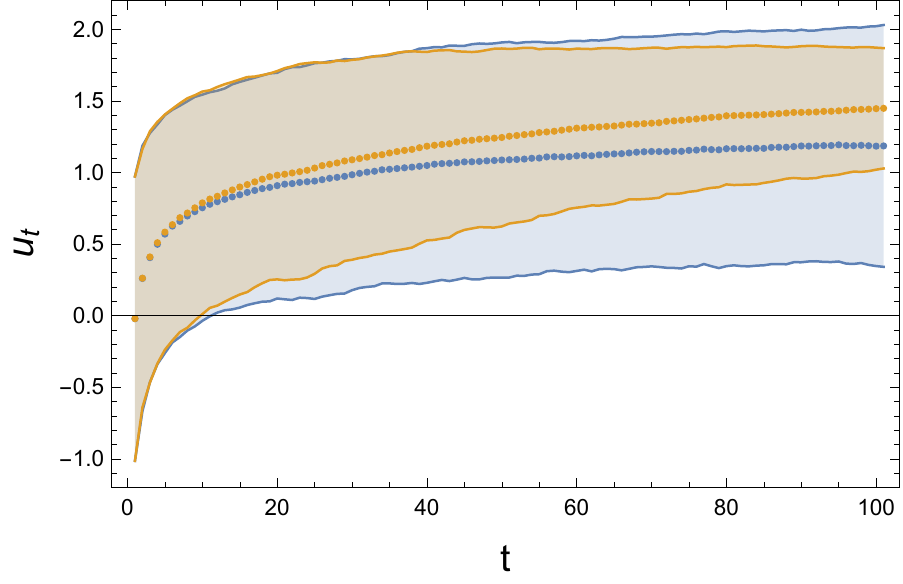}
  \caption{Gaussian model $\mathcal{N}(0,1)$}
  \label{utumtGn10000s1l1}
\end{subfigure}%
\begin{subfigure}{.49\textwidth}
  \centering
  \includegraphics[width=\linewidth]{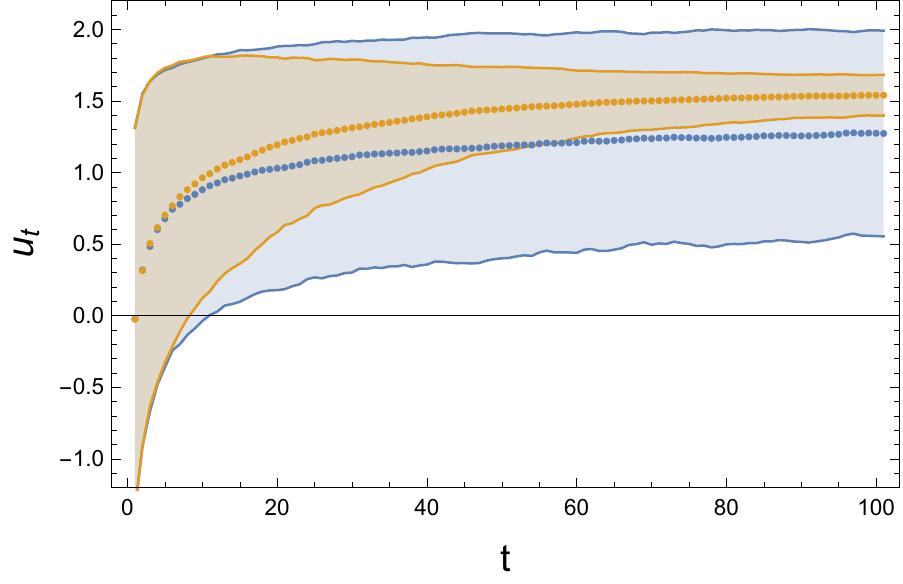}
  \caption{Uniform model $U(-2,2)$}
  \label{utumtUN10000s1l1}
\end{subfigure}
\caption{Average utility evolution in the the system with marriage and $\Lambda=1$ (orange) and without marriage (blue) in the Gaussian and Uniform models with $N=10.000$.}
\label{averageutilityandvariance}
\end{figure}

It is also interesting to plot the evolution of $u_t$ with the variance as we did in the previous section for the model with and without marriage. In Fig. (\ref{averageutilityandvariance}), we see that the system with marriage and $\Lambda=\sigma$, have a higher mean and a lower variance. If agents value equality with respect to relationships, this fact can be interpreted as another benefit of marriage. 

It is important to emphasize that the benefits of marriage as a search strategy disclosed using this model may be irrelevant when other aspects of marriage are considered. There are many layers of our preferences concerning relationships that are not taken into account by this model. For example, although an agent that does not marry will end up on average with someone with a lower affinity, it nevertheless may be the case that this agent gets a surplus of utility associated with the fact of participating in multiple relations during his lifetime. As another example, in general it is painful to break apart. However, this loss is not considered in the model. But would rather add another benefit to the society with marriage. Overall, evidence seems to suggest that marriage is beneficial in a number of respects \cite{doi:10.2307/2061670}.

A salient simplification of our model is that in real life, the affinity between agents is not time-independent. It is worth to mention, that if we include a time-dependent affinity, then our model will already include divorce. Indeed, given a married couple, their affinity may get smaller than $\Lambda$. When this happens, the married couple will unmarry and may break apart if one of the participants finds another mate. A time-dependent affinity would also account for the spontaneous break-up of couples. It suffices that the affinity between the couple gets smaller than the utility for one of the mates to be single. Understanding the average dynamical behavior of the affinity between agents is beyond the scope of this paper. Perhaps, the affinity is an oscillatory function with higher amplitude as the time with a given mate passes. As it often happens, old couples frequently oscillate between high-affinity and low-affinity states.

\subsection{Discussion, comments and empirical comparison}
\label{discussionandcomments}

A possible point of concern, is the fact that we did not distinguished man and woman as in the original marriage problem \cite{10.2307/2312726}. One could then think that we are actually dealing with something related to the roommate problem instead \cite{10.2307/2312726}. This distinctions, however, are not important for our model. We can partition society into two, making matching just between agents not of the same group. The evolution of average utility for such a system is visually indistinguishable from Fig. (\ref{utumtGn10000s1l1}) and Fig. (\ref{utumtUN10000s1l1}), giving the same pattern and disclosing the same benefits outlined before. One can even partition society into groups of different sizes, and the results are still very similar. For this situation, the matching in Eq. (\ref{matching}) is a random surjective function from the smallest partition to the largest one. A random subset of agents that make the large partition is left out from the matching process at each step. Several variations are possible in this setup. For example, we could select a subset of one of the partitions and make it match with itself as well. This subset could then be interpreted as bisexual agents. This flexibility of our model goes in line with the present diversity of modern relationships and contrasts with results for the matching problem. Many results, such as the Gale-Shapley algorithm, depend on the fact that both sets in the matching process have the same number of elements \cite{roth1992two}.

It is interesting to compare the value for the average utility attained in our model with the value obtained using the Gale–Shapley algorithmic solution to the matching problem \cite{10.2307/2312726}. We use the Gale-Shapley algorithm to find a stable matching and then we computed the average utility associated with it $u_{G-S}$. We did this for different values of $N$. For all $N$ studied, $u_{G-S}$ is much higher than the values obtained in our model. For example, in the Gaussian model with $N\simeq 1.000$ we found $u_{G-S}\simeq 4$. This is another indication that the concept of stable matching is not relevant to understand the matching of couples in large societies. However, the fact that $u_{G-S}$ is larger than $u_t$ for $t\simeq100$, implies that the market of relationships in large societies is sub-optimal. If in the future we are able to infer the affinity between agents using correlated manifest properties, such as age, personality, political bias, level of education, etc., then we could use the Gale-Shapley algorithm to spell a stable matching, and society average utility with respect to relationships would be much higher.

\begin{figure}
\centering
\begin{subfigure}{.49\textwidth}
  \centering
  \includegraphics[width=\linewidth]{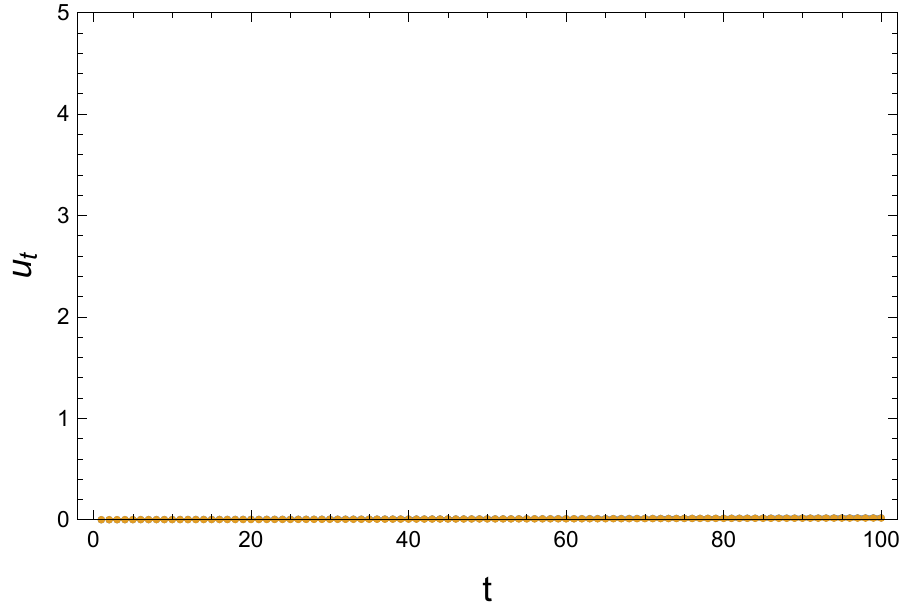}
  \caption{$\mu=-2$}
  \label{mum2}
\end{subfigure}%
\begin{subfigure}{.49\textwidth}
  \centering
  \includegraphics[width=\linewidth]{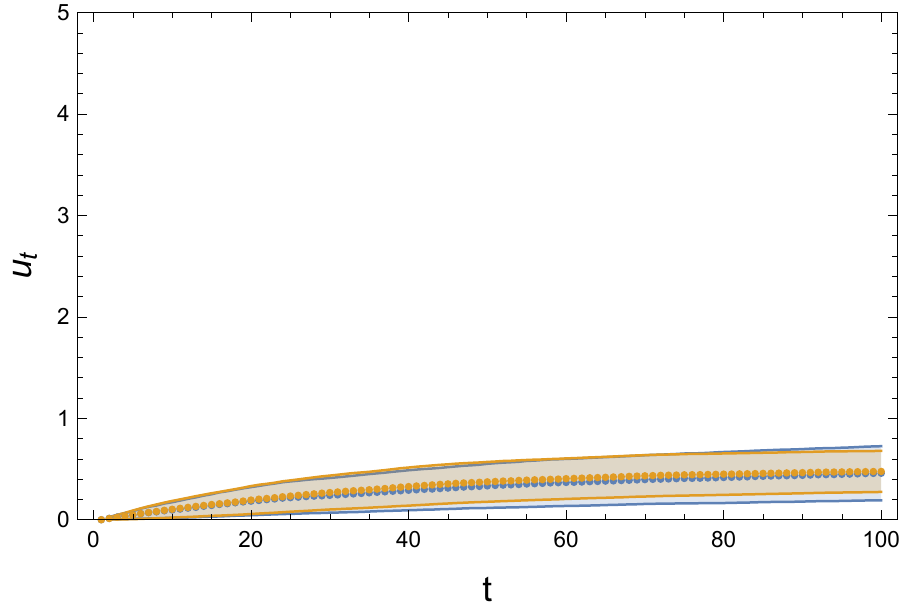}
  \caption{$\mu=-1$}
\end{subfigure}\vspace{5mm}
\begin{subfigure}{.49\textwidth}
  \centering
  \includegraphics[width=\linewidth]{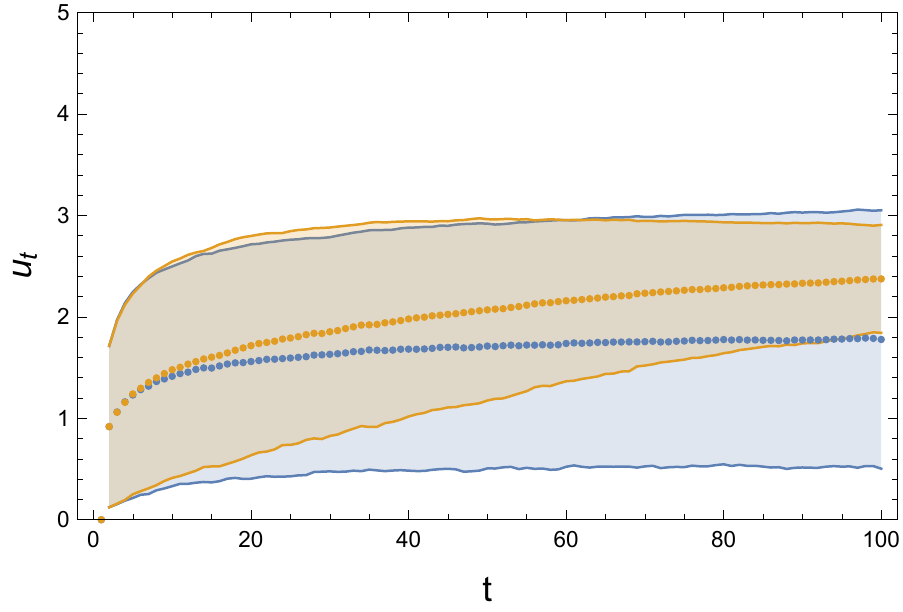}
  \caption{$\mu=1$}
\end{subfigure}%
\begin{subfigure}{.49\textwidth}
  \centering
  \includegraphics[width=\linewidth]{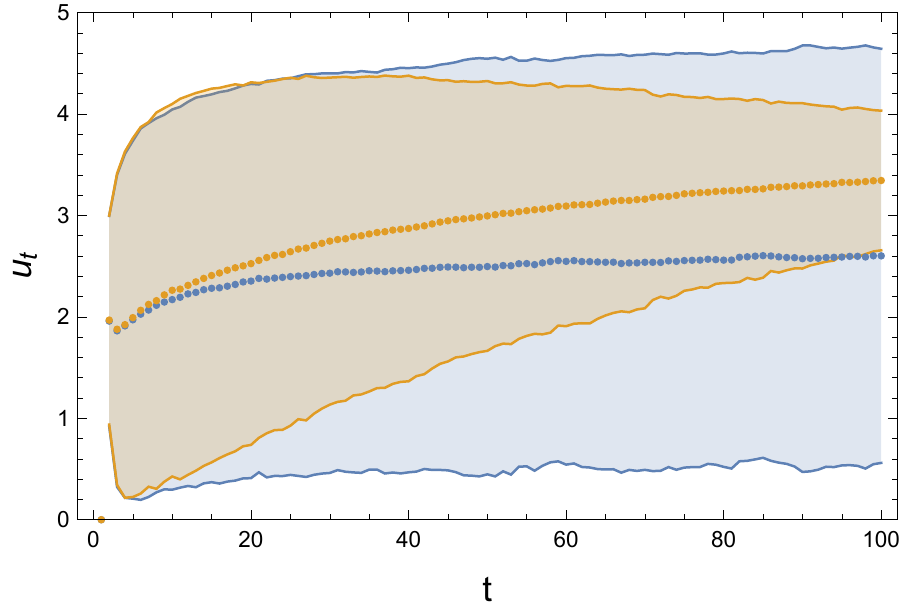}
  \caption{$\mu=2$}
  \label{mu2}
\end{subfigure}
\caption{Average utility evolution in the model with $A_{kk}=0$ for all $1\leq k\leq N$ and $A_{ij}\sim\mathcal{N}(\mu,1)$ for all $1\leq i\leq N$ and $1\leq j\leq N$ with $i\neq j$ and $N=10.000$. The orange dots corresponds to the system with marriage and $\Lambda=\mu+1$, and the blue dots to the system without marriage.}
\label{compchangingmu}
\end{figure}

As mentioned in the discussion after Fig. (\ref{InitialHistograms}), we assumed for concreteness that the diagonal and off-diagonal elements of the affinity matrix are given by the same distribution. So, one might rightly worry that our results are not valid if this is not the case. To address this concern, we simulate our model assuming that $A_{kk}=0$ for all $1\leq k\leq N$ and that $A_{ij}=\mathcal{N}(\mu,1)$ for all $1\leq i\leq N$ and $1\leq j\leq N$ with $i\neq j$, for different values of $\mu$ and $\Lambda=\mu+\sigma=\mu+1$. Note that $A_{kk}=0$ is obtained formally using $\mu=0$ and taking the limit $\sigma\rightarrow0$ for the distribution $\mathcal{N}(\mu,\sigma)$ that generates the diagonal elements of the affinity matrix. Furthermore, in this limit, there is no loss of generality in assuming that $\sigma=1$ for the off-diagonal elements. This happens because, when $A_{kk}=0$ for all $1\leq k\leq N$, the dynamical equations (\ref{singleevolution}) and (\ref{coupleevolution}), are still invariant under re-scaling of $A_{ij}$ and therefore the results obtained for $\sigma=1$ can be translated to generic $\sigma$ as we did in (\ref{generalmusigma}). The results we obtained are presented in Fig. (\ref{compchangingmu}). The following considerations are helpful to interpolate the figures and get a better picture of the system as a function of $\mu$. For values of $\mu$ less than $-2\sigma=-2$, an agent is unlikely to find one it likes more than being single, and the behavior of the system will be very similar to that of Fig. (\ref{mum2}). On the other extreme, if $\mu>2\sigma=2$, agents will prefer to date any other agent than being single as they did for $\mu=2\sigma=2$, and the dynamics will be very similar to that of Fig. (\ref{mu2}). The difference, in this case, will be that the average utility will range on a different scale set by $\mu$. Because for all values we have the average utility in the system with marriage higher than in the system without it, we are confident in concluding that this is a robust feature of our model. Therefore, the advice we extracted seems to hold in more general contexts.

Another objection to the results presented in the previous section is that the assumption of a homogeneous society is clearly false. What is interesting, however, is that for a finite number of agents, this homogeneity is already broken. More precisely, for all agents $k$, its affinity with the other agents is given by $\{A_{jk}:1\leq j\leq N\textrm{ and } j\neq k\}$, which is generated by the same distribution but is, in general, a different collection of numbers for each agent. To make this point more clear consider the following. We can select all agents $k$ such that the average of the collection $\{A_{jk}:1\leq j\leq N\textrm{ and } j\neq k\}$ is larger than a certain positive value. We can pick this value in a way that we have just $1\%$ of all agents selected. This will represent the $1\%$ most liked agents of the system. Conversely, we can do the same and get the $1\%$ less liked agents. It is then interesting to follow the evolution of average utility for these different sets of agents.
One finds that marriage is more beneficial to those that are not that liked. This agrees with intuition. If everyone likes you, you should not be concerned with marriage as a search strategy. It is easier for you to find a nice partner that also likes you and your mate is less likely to leave you because of someone else.

It is hard to measure the average utility in a real societies, but the number of married couples is not. By comparing the share of married couples in our model and existent societies we can test the adequacy of our model to describe reality. We have

\begin{figure}[H]
    \centering
    \begin{minipage}{0.47\textwidth}
        \centering
        \includegraphics[width=\textwidth]{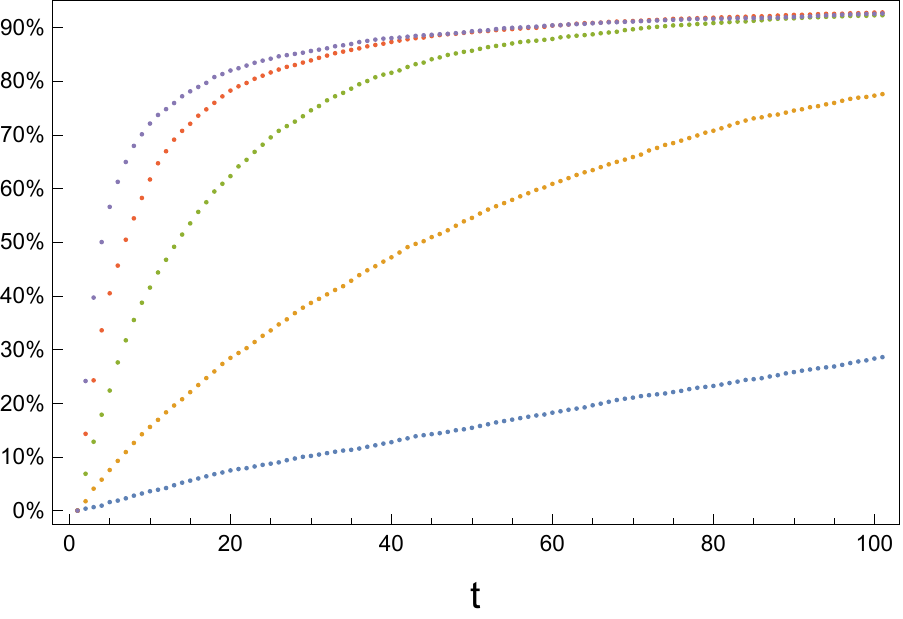}
        \caption{Share of married agents in the Gaussian model $\mathcal{N}(0,1)$ with $N=10.000$ at each step for different values of $\Lambda$. From the bottom to the upper lines we have $\Lambda=1.5,1,0.5,0,-2$.}
        \label{mcsimulation}
    \end{minipage}\hfill
    \begin{minipage}{0.47\textwidth}
        \centering
        \includegraphics[width=\textwidth]{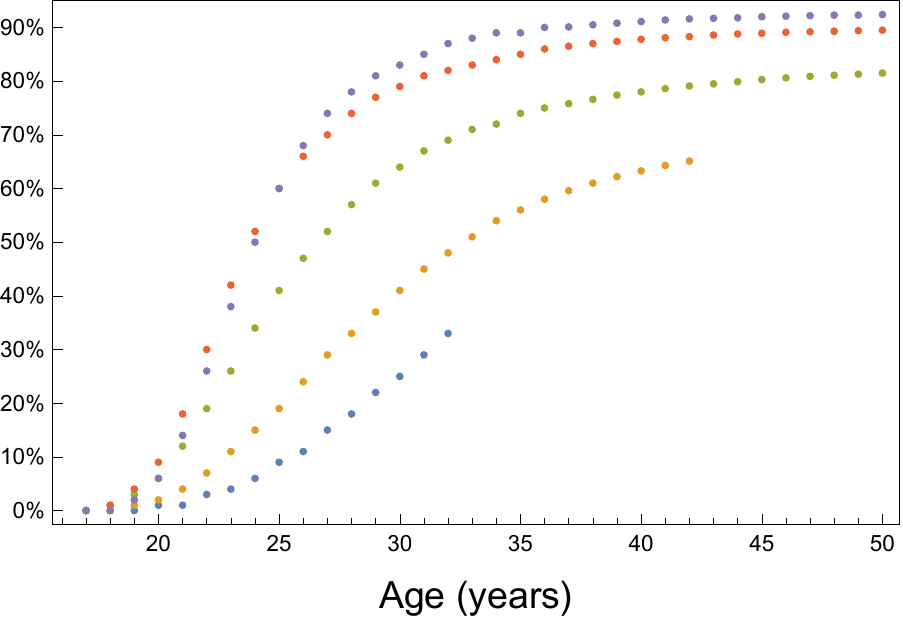}
        \caption{Share of men in England and Wales who were married by a certain age \cite{owidmarriagesanddivorces}. From the bottom to the upper lines we have men who were born at $1980,1970,1960,1950,1940$ respectively.}
        \label{mcrealdata}
    \end{minipage}
\end{figure}

\noindent which is arguably a good agreement when taken into account that we purposely ignore some relevant aspects of marriage, such as divorce. We highlight, though, that we could implement divorce by having a time-dependent affinity, as mentioned in the last paragraph of section (\ref{benefitsofmarriage}). We also call attention to the fact that the similar behavior of the two plots was not built-in our stochastic matching model in any way. It is a pure dynamical consequence. 

The trends of a declining number of married men in Fig. (\ref{mcrealdata}) can be interpreted in the framework of our model as a shift to higher $\Lambda$. This agrees with rough expectations. In the past, marriage was enforced by tradition and many couples get married without high affinity. Progressive movements diminished the forces of tradition and people became more rigorous to get married. If we go further and assume that each generation in Fig. (\ref{mcrealdata}) is associated with a value of $\Lambda$ with the same color in Fig. (\ref{mcsimulation}), we can see in Fig. (\ref{lcompUG10000s1}) that the average utility concerning relationships increased from 1940 to 1970, and perhaps starts decreasing from 1970 to 1980. This being true, would suggest that the progressive movements benefited the matching markets of traditional societies by increasing $\Lambda$. But perhaps, $\Lambda$ went too high and needs to be balanced. The right balance can be achieved following the piece of advice extracted from the results of the previous section: get married if you like your partner more than a sigma above average. As mentioned before, the steps $t$ are likely not linearly related to chronological time. If initial steps are compressed, as it is reasonable to assume, the initial convex behavior of the real-world data in Fig. (\ref{mcrealdata}) may be attained. 

To end this section, we will give an operational answer to the question that opened this work. The affinity matrix contains information that is not accessible to the agents. They discover their affinity just when they match. Therefore, agents don't know the value of sigma before the evolution of the system. However, one can use the utility attained from previous matchings as an input to calculate a mean and a variation and then use these estimates to decide if one proposes marriage or not. More precisely, let $\{u_t:0\leq t\leq T\}$ be the utility of some agent at steps $0\leq t\leq T$. This agent should propose marriage at step $T+1$ if:

\begin{align}
    u_{T+1}>\frac{1}{T}\sum_{t=0}^Tu_t+\sqrt{\Big(\frac{1}{T}\sum_{t=0}^Tu_t^2\Big)-\Big(\frac{1}{T}\sum_{t=0}^Tu_t\Big)^2}.
\end{align}

\noindent It is left to the agent the inner task of correctly estimate its affinity with the agents it already matched. This is by no mean a simple task, as we so commonly fool ourselves in matters of love. Perhaps, this is not even a well defined quantity after all.

\section{Analytical formulation in the large \texorpdfstring{$N$}{N} limit}
\label{analytical} 

\begin{figure}
\begin{subfigure}{.48\textwidth}
  \centering
  \includegraphics[width=\textwidth]{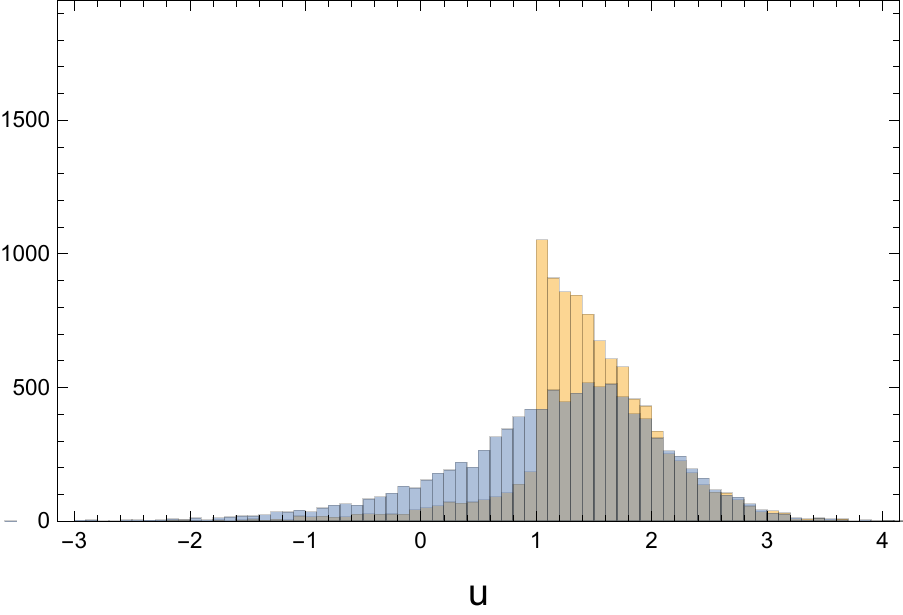}
    \caption{Gaussian model, $U_0\sim\mathcal{N}(0,1)$.}
  \label{averageutilityGaussian}
\end{subfigure}%
\hspace{5mm}
\begin{subfigure}{.48\textwidth}
  \centering
  \includegraphics[width=\textwidth]{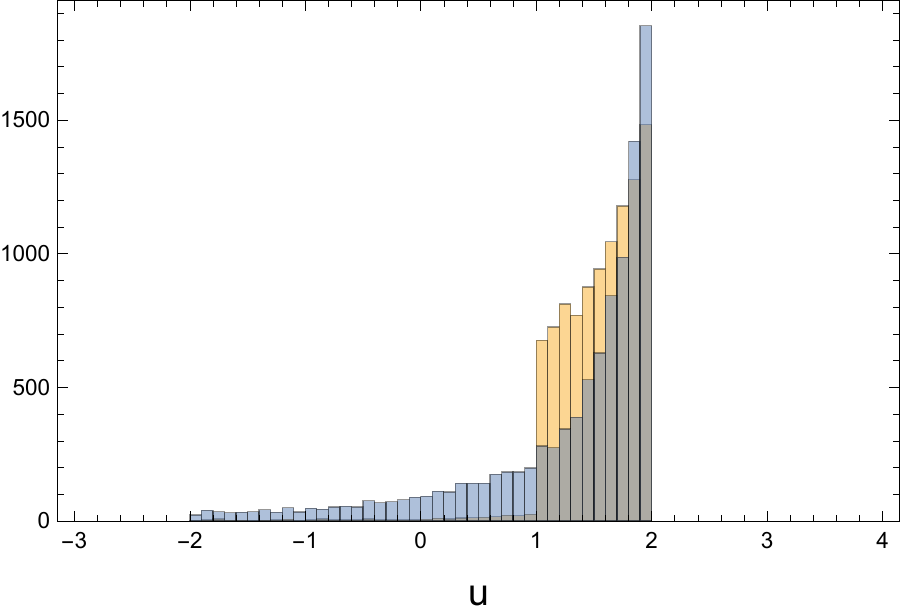}
  \caption{Uniform model, $U_0\sim U(-2,2)$.}
  \label{averageutilityNormal}
\end{subfigure}
\caption{Histogram of $\{u_t^k:1\leq k\leq 10.000\}$ at $t=100$ with bin size $0.1$ for the model with marriage with $\Lambda=1$ (orange) and without marriage (blue). The vertical axis is the number of agents and the horizontal axis the utility. Together with the histograms in Fig. (\ref{InitialHistograms}), this gives a visualization of the evolution of the system.}
\end{figure}

In the previous section, we focused on the evolution of the average utility and its variance, disclosing some of the benefits of marriage as a search strategy. The average utility and its variance are aggregate quantities calculated from the set of values $\{u_t^k:1\leq k\leq N\}$ at each step $t$. However, this set of values contains much more information as one can see visually with its histograms in Fig. (\ref{averageutilityGaussian}) and Fig. (\ref{averageutilityNormal}). Another way to collect this information is to introduce a random variable $U_t$ with a finite number of outcomes $\{u_t^k:1\leq k\leq N\}$ occurring with probabilities $\frac{1}{N}$. Then, the average utility will be given by the expectation value of $U_t$, i.e. $E[U_t]$, and the variance by $(E[U_t^2]-E[U_t]^2)^{\frac{1}{2}}$, where

\begin{align}
    E[U^n_t]\equiv\frac{1}{N}\sum_{i=1}^N(u_t^k)^n,
\end{align}

\noindent is the $n-$moment of the random variable $U_t$. This form of collecting the information shows more precisely what information we are missing by analyzing just the average utility and its variance: we are missing the higher moments of $U_t$.

At $t=0$ we have $u_{0}^k=A_{kk}\sim U_0$ which is given by a known distribution with probability density function $p_0(u)$. Therefore, $U_0$ is a random variable sampled by this distribution and approximates the actual distribution as $N\rightarrow\infty$. In this limit, we have

\begin{align}
    E[U_0^n]=\lim_{N\rightarrow\infty} \frac{1}{N}\sum_{k=1}^NA_{kk}^n= \int_{-\infty}^{\infty} u^np_0(u)du.
\end{align}

\noindent We see that we have complete information of the system at $t=0$ as $N\rightarrow\infty$ because we know the probability distribution of $U_0$ and we can calculate explicit all moments $E[U^n_0]$. For finite $N$, if we run the simulation and compare $E[U_0^n]$ with the corresponding quantity in the limit as $N\rightarrow\infty$, they will not be exactly equal. However, if we do a large number of simulations and average the results we will get better approximations to the value obtained in the limit. This considerations shows that the probability distribution $p_t(u)$ that samples the values $\{u_t^k:1\leq k\leq N\}$ is more fundamental than the actual values. 

The objects of interest in this section are the probability distribution function $p_t(u)$ and the moments $E[U_t^n]$. The first gives the probability of finding an agent with utility $u$ at time $t$, and the second characterizes this distribution. In fact, both objects can be thought as different aspects of the same mathematical construct seems one can be built from the other. To get $E[U_t^n]$ from $p_t(u)$ it suffices to integrate $u^n$ with weight $p_t(u)$. To get $p_t(u)$ from $E[U_t^n]$ is a bit more laborious, one needs to compute the characteristic function

\begin{align}
    E[e^{ik U_t}]=\int_{-\infty}^{\infty} e^{ik U_t}p_t(u)=\sum_{n=0}^\infty E[U_t^n]\frac{(ik)^n}{n!},
    \label{characteristicfunction}
\end{align}

\noindent using the moments $E[U_t^n]$, and then take the inverse Fourier transform. From this discussion, one sees that a formula for computing $E[U_{t}^n]$ from $E[U_{t-1}^n]$ is equivalent to finding $p_t(u)$ from $p_{t-1}(u)$. This is the task of this section. In section (\ref{firststep}) we will work out $p_1(u)$. In section (\ref{nextstep}) we will write the general evolution equation for $p_t(u)$ in terms of its moments $E[U_t^n]$. In both sections we will be concerned with our model without marriage. Then, in section (\ref{marriedcouplesdistribution}) we will find the distribution for married couples.

Before starting, we should mention that, although more tractable analytically and more fundamental seems approximated by any finite implementation, this continuous version of the problem lack some features that approximate the computational implementation with reality. Apart from the salient unrealistic fact that real societies have a finite number of agents, another important feature of $N\rightarrow\infty$ is that in this limit the society is indeed homogeneous. For a finite number of agents $N$, homogeneity is broken and we end up with agents that are more or less liked than others 
. This departure from homogeneity among agents is certainly an important feature of real societies as we all have experiences with people that are more liked than others. We explored this feature in the previous section, and we mentioned that the more liked agents receive, on average, a smaller benefit from marriage than the agents who are not that liked. This sort of analysis is not possible in the continuum version where the system is indeed homogeneous.

\subsection{Warm-up: distribution after the first step}
\label{firststep}

As a preparation for the next section, we will find an analytical equation for $E[U_1^n]$ for our model without marriage given the random variable $U_0$. To do that, note that agents are either coupled or singles at each step. Couples and singles behave differently and we need to consider them separately to find the equation for $E[U_1^n]$. Furthermore, because agents are either coupled or singles, $p_t(u)$ can be decomposed as

\begin{align}
    p_t(u)=b_tp_{ct}(u)+r_tp_{st}(u),
    \label{decomposition}
\end{align}

\noindent with $p_{ct}(u)$ and $p_{st}(u)$ respectively the utility distribution for the coupled and single agents, and $b_t,r_t$ the proportion of couples and singles in the population. If we introduce random variables for the couples and singles, $C_t$ and $S_t$, the decomposition (\ref{decomposition}) is equivalent to the following relation between the moments of these random variables:

\begin{align}
    E[U_t^n]=b_{t}E[C_t^n]+r_tE[S_t^n],
    \label{UtCtSt}
\end{align}

\noindent therefore, to find $E[U_1^n]$ we need to find $b_1$, $E[C_1^n]$, $r_1$ and $E[S_1^n]$.

At $t=0$ all agents are singles therefore $b_0=0$ and $r_0=1$. The diagonal elements of the affinity matrix encodes the utility when agents are singles and is an input of the model. Hence, these elements are generated by the random variable $U_0$ with probability distribution equals $p_0(t)$. The off-diagonal elements of the affinity matrix may be generated by another distribution as emphasized in the discussion after Fig. (\ref{InitialHistograms}). We will denote the probability distribution that generates these elements by $p_A(u)$.

Let's start with $b_1$. This number is the probability of forming a couple at $t=1$. All agents are singles at $t=0$ and match at random at $t=1$. If two agents, described by the random variables $K$ and $L$ match, they will form a couple if they like each other. More precisely,

\begin{align}
    c_0(K,L)\equiv\int_K^\infty p_A(u)du\int_L^\infty p_A(u)du,
    \label{c0}
\end{align}

\noindent where $p_A(u)$ is the probability distribution function that generates the off-diagonal elements of the affinity matrix.  The random variables $K$ and $L$ are distributed according to $U_0$ seems agents are single at $t=0$. Then, the expectation value of $c_0(K,L)$ gives $b_1$, that is

\begin{align}
    b_1=E[c_0(K,L)]\equiv\int_{-\infty}^\infty\int_{-\infty}^\infty c_0(k,l)p_0(k)p_0(l)dkdl,\qquad A,B\sim U_0.
    \label{b1}
\end{align}

Conversely, the probability that agent $K$ will stay single at $t=1$, is the complement of $c_0(K,L)$. This quantity is composed of two contributions, either agent $L$ likes agent $K$ but agent $K$ does not like agent $L$. Or agent $L$ does not like agent $K$. More precisely,

\begin{align}
    s_0(K,L)\equiv1-c_0(K,L)=\int_{-\infty}^Kp_A(u)du\int_L^\infty p_A(u)du+\int_{-\infty}^Lp_A(u)du.
    \label{s0}
\end{align}

\noindent As before,

\begin{align}
    r_1=E[s_0(K,L)],\qquad K,L\sim U_0.
    \label{r1}
\end{align}

Because (\ref{s0}) is the complement of (\ref{c0}), it is easy to check that indeed

\begin{align}
    E[U_1^0]\equiv b_1E[C_1^0]+r_1E[S_1^0]=b_1+r_1\equiv E[c_0(A,B)+s_0(A,B)]=E[1]=1.
\end{align}

\noindent For the homogeneous Gaussian and Uniform model we studied, with all elements of the affinity matrix generated by the same distribution, we have

\begin{align}
    b_1=\frac{1}{4},\qquad r_1=\frac{3}{4}.
    \label{b1r1}
\end{align}

\noindent This results can be compared with simulations by counting the number of couples and singles at $t=1$ and dividing by $N$. In a typical simulation of the Uniform model with $N=10.000$ and $\sigma=1$ we got

\begin{align}
    b_1^{\textrm{exp}}=\frac{2430}{10000}=24.3\%,\quad r_1^{\textrm{exp}}=\frac{7570}{10000}=75.7\%,
\end{align}

\noindent which differs from the $N\rightarrow\infty$ values, in the $0.1\%$ level.

Now, it is easy to generalize the expressions for $b_1$ and $r_1$ to give $E[C^n_1]$ and $E[S_t^n]$ instead. We have $E[C_1^0]=E[S_1^0]=1$, and for $n>0$ we have

\begin{align}
    E[C_1^n]=E[c_n(K,L)],\qquad E[S_1^n]=E[s_n(K,L)],\qquad K,L\sim U_0,
    \label{C1nS1n}
\end{align}

\noindent where

\begin{align}
    & c_n(K,L)\equiv\int_K^\infty u^np_A(u)du\int_L^\infty p_A(u)du,\label{cnsingle}\\
    & s_n(K,L)\equiv 1-c_n(K,L)= K^n\int_{-\infty}^Kp_A(u)du\int_L^\infty p_A(u)du+K^n\int_{-\infty}^Lp_A(u)du.\label{snsingle}
\end{align}

\noindent Using Eq. (\ref{b1}), (\ref{r1}) and (\ref{C1nS1n}) in Eq. (\ref{UtCtSt}) we can compute $E[U_1^n]$ as we wanted.

For the Uniform model with the diagonal and off-diagonal elements of the affinity matrix generated by the same distribution we have 

\begin{align}
    p_0(u)=p_A(u)=\frac{1}{4\sigma}\begin{cases} 1 & \textrm{if }-2\sigma\leq u\leq 2\sigma \\ 0 
    & \textrm{otherwise}\end{cases}.
    \label{p0}
\end{align}

\noindent Using this formula in Eq. (\ref{C1nS1n}) we can compute

\begin{align}
    E[C_1^n]=\frac{(2 n+3) (2\sigma)^n+(-2\sigma)^n}{2(n+1) (n+2)},\qquad E[S_1^n]=\frac{(4 n+7) (2\sigma)^n+(2 n+5) (-2\sigma)^n}{2 (n+1) (n+2)},\label{C1nS1nuniform}
\end{align}

\noindent which together with Eq. (\ref{b1r1}) can be used to find $E[U_1^n]$. By considering the couples and singles separately at step 1, we can compare the values of Eq. (\ref{C1nS1nuniform}) with the sample moments obtained in the simulation. This comparison gave good agreement, although we were not systematic in estimating the uncertainties. By taking the inverse Fourier transform of the characteristic function given by Eq. (\ref{characteristicfunction}), we can compute the utility probability distribution of couples and singles, 

\begin{align}
    & p_{c1}(u)=\frac{2\sigma+u}{2\sigma}p_0(u),\label{pc1}\\
    & p_{s1}(u)=\frac{6\sigma+u}{6\sigma}p_0(u)\label{ps1}.
\end{align}

\noindent Their sum, with factors $b_1$ and $r_1$ gives the probability distribution of $U_1$ namely

\begin{align}
    p_1(u)=b_1p_{c1}(u)+r_1p_{s1}(u)=\frac{4\sigma+u}{4\sigma}p_0(u).\label{p1}
\end{align}

\noindent A visual check of the results is to compare the normalized histogram obtained in the simulations and the plot of these probability distribution functions. We have

\begin{figure}[H]
\begin{subfigure}{.5\textwidth}
  \centering
  \includegraphics[width=\textwidth]{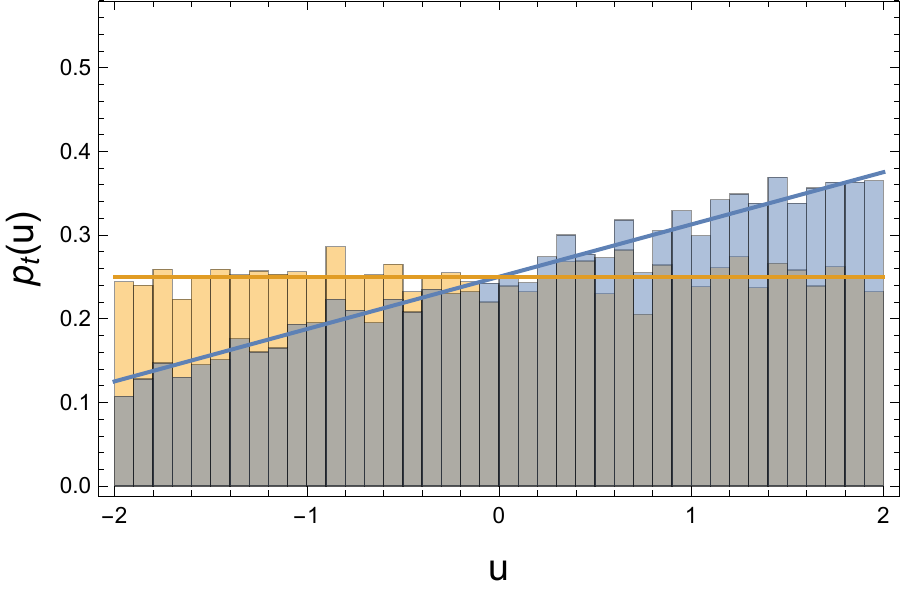}
  \caption{Distributions at step 0 and 1.}
  \label{p0p1}
\end{subfigure}%
\begin{subfigure}{.5\textwidth}
  \centering
  \includegraphics[width=\textwidth]{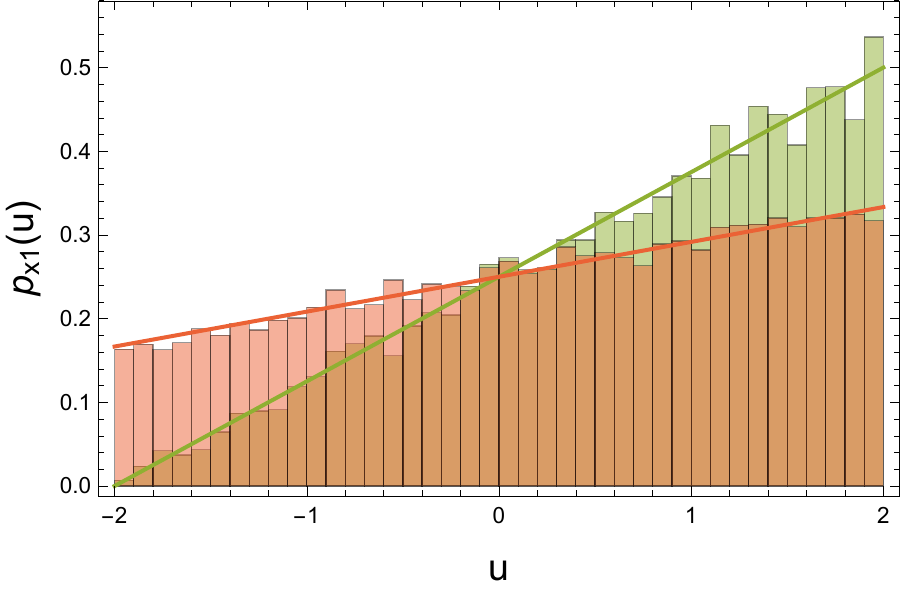}
  \caption{Distributions for singles and couples at 1}
  \label{pc1ps1}
\end{subfigure}
\caption{In Fig. (\ref{p0p1}), the orange line is the initial distribution Eq. (\ref{p0}) and the blue line the distribution at step 1 Eq. (\ref{p1}) with $\sigma=1$. In Fig. (\ref{pc1ps1}), the orange and green lines are respectively the distribution for couples Eq. (\ref{pc1}) and singles Eq. (\ref{ps1}) at step $1$, both for $\sigma=1$. We also plotted the corresponding histogram for the simulation with $N=10.000$ and $\sigma=1$.}
\end{figure}

For the Gaussian system, the integrals are more involved, but it would be interesting to explore and see if it is also possible to find an expression for $p_{c1}(u)$ and $p_{s1}(u)$ in terms of familiar functions. As the first step in this direction, we calculate the mean of $U_1$ in the Gaussian model. With Eq. (\ref{p1}), one can also calculate it for the Uniform model. These two results are:

\begin{align}
    E[U_1]=\begin{dcases} \frac{\sigma}{2\sqrt{\pi}}, & \textrm{if Gaussian,}\\ \frac{\sigma}{3}, & \textrm{if Uniform.}
    \end{dcases}
\end{align}

\noindent It is nice to see that both are consistent with the dictionary explained in Eq. (\ref{generalmusigma}).

It is currently beyond our mathematical prowess to prove that the distribution $p_1(u)$ is indeed what is obtained in the simulation at step $1$ in the limit $N\rightarrow\infty$. However, the intuitive derivation together with the visual fit between the histograms and the distributions are already compelling reasons to believe that this is indeed the case. In addition, one can compare the sampled moments of the simulation with the moments of the distribution. Such an analysis also gives a good agreement between the two. In any case, it would be interesting to have formal proof of this fact.

\subsection{Evolution equations in the large \texorpdfstring{$N$}{N} limit}
\label{nextstep}

In the previous section, the equations were simple because all agents were single at $t=0$. Now, we have a distribution $p_{c1}(u)$ and $p_{s1}(u)$ for the utility of couples and singles. The singles will behave as in the previous section, but the couples behave differently.

If an agent $K$ and $M$ are a couple at step $1$ and they match with agents $L$ and $N$ at step $2$, then the probability for agent $K$ to be in a couple at step $2$ decomposes in five pieces. If agent $L$ likes agent $K$, either agent $K$ also likes agent $L$ and they form a couple or agent $K$ does not like agent $L$ but agents $M$ and $N$ does not form a couple as well. Otherwise, if agent $L$ does not like agent $K$, then it will remain in a couple if agents $M$ and $N $ do not form a couple. Furthermore, the probability that agents $M$ and $N$ will not form a couple can be decomposed into two parts. Either $M$ likes $N$ but $N$ does not like $M$, or $N$ does not like $M$. Already making the generalization we did from Eq. (\ref{c0}) to Eq. (\ref{cnsingle}) we will have

\begin{align}
    \begin{split}
    c_n(K,L,M,N)\equiv & \int_{L}^\infty p_A(u)du\Big(\int_{K}^\infty p_A(u)u^ndu+f(K,M,N)\int_{-\infty}^{K}p_A(u)du\Big)\\
    & +f(K,M,N)\int_{-\infty}^{L} p_A(u)du,
    \end{split}
    \label{cncouple}
\end{align}

\noindent with 

\begin{align}
    f(K,M,N)\equiv K^n\int_{N}^\infty p_A(u)du\int_{-\infty}^{M}p_A(u)du+ K^n\int_{-\infty}^{N}p_A(u)du.
\end{align}

\noindent The moments of $C_{t}$ at each time will have a contribution that comes from the singles given by Eq. (\ref{cnsingle}) and a part given by Eq. (\ref{cncouple}). Each part is proportional to the fraction of singles and couples in society at the previous step. We hence have

\begin{align}
    b_{t+1} & =b_{t}E[c_0(K_c,L,M,N)]+r_tE[c_0(K_s,L)],\label{evolbtn}\\
    E[C_{t+1}^n] & =b_{t}E[c_n(K_c,L,M,N)]+r_tE[c_n(K_s,L)],\qquad n>0\label{evolECtn}
\end{align}

\noindent with $K_s\sim S_t$, $K_c,M\sim C_t$, $L,N\sim U_t$. In words, $K_s$ is a single, $K_c,M$ a couple, and $L,N$ either couple or single. Note that for $n=0$, $E[C_{t+1}^n]$ is not defined by Eq. (\ref{evolECtn}), seems $E[C_{t+1}^0]=E[1]=1$ and not $E[C_{t+1}^0]=b_{t+1}\neq1$.

Conversely, agent $K$ will be single at $t=2$ if $L$ and $N$ form a new couple, and either if $L$ does not like $K$ or if $L$ likes $K$ but $K$ does not like $L$. More precisely, and generalizing we have

\begin{align}
    \begin{split}
    s_n(K,L,M,N)\equiv 1-c(K,L,M,N) & =g(M,N)\int_{L}^\infty p_A(u)du\int_{-\infty}^{K}p_A(u)du\\
    & +g(M,N)\int_{-\infty}^{L} p_A(u)du,
    \end{split}    
    \label{coupled}
\end{align}

\noindent with

\begin{align}
    g(M,N)\equiv E[U_0^n]\int_{M}^\infty p_A(u)du\int_{N}^\infty p_A(u)du,
\end{align}

\noindent where we used $E[U_0^n]$ instead of $K$ to account for the fact that in this case, the agent $K$ will end up single getting back to its initial distribution $U_0$. Because $U_0$ does not appear anywhere else, we already use its expectation value instead of $U_0$. Taking the expectation values as we did in Eq. (\ref{evolECtn}) and Eq. (\ref{evolbtn}) gives

\begin{align}
    r_{t+1} & =b_{t}E[s_0(K_c,L,M,N)]+r_tE[s_0(K_s,L)],\label{evolrtn}\\
    E[S_{t+1}^n] & =b_{t}E[s_n(K_c,L,M,N)]+r_tE[s_n(K_s,L)],\qquad n>0\label{evolEStn}
\end{align}

\noindent and conversely $K_s\sim S_t$, $K_c,M\sim C_t$, $L,N\sim U_t$. As before, for $n=0$, $E[S_{t+1}^n]$ is not defined by Eq. (\ref{evolECtn}), seems $E[S_{t+1}^0]=E[1]=1$ and not $E[S_{t+1}^0]=r_{t+1}\neq1$.

Note that the evolution equations (\ref{evolbtn}), (\ref{evolECtn}), (\ref{evolrtn}) and (\ref{evolEStn}) are equal the ones given in the previous section because for $t=0$ we have $b_0=0$ and $r_0=1$ which sets the first terms in each case to zero. In addition, at $t=0$ we have $S_0\sim U_0$. It is easy to prove that this evolution equation preserves probability. Because

\begin{align}
    s_0(K,L,M,N)+c_0(K,L,M,N)=s_0(K,L)+c_0(K,L)=1,
\end{align}

\noindent we have

\begin{align}
    b_{t+1}+r_{t+1}=b_t+r_t,
\end{align}

\noindent therefore, if $E[U_0^0]=b_0+r_0=1$, by induction we will have $E[U_t^0]=b_t+r_t=1$ for all $t$. Combining all this expressions we had accomplished the desired result of formulating the dynamics of the system in an analytic way. We built a mathematical machinery to compute $p_{t+1}(u)$ given the four components of $p_t(u)=b_tp_{ct}(u)+r_tp_{st}(u)$.

Using the trick of computing the Fourier transform of the characteristic function given in Eq. (\ref{characteristicfunction}), for the Uniform model at $t=2$ we have

\begin{align}
    p_2(u)=\frac{1861}{5184}p_{c2}(u)+\frac{3323}{5184}p_{s2}(u),
    \label{p2}
\end{align}

\noindent with 

\begin{align}
    p_{c2}(u)=\frac{3 (2\sigma+u) (2258\sigma+335 u)}{7444(2\sigma)^3},\qquad p_{s2}(u)=\frac{3233 (2\sigma)^2+3672\sigma u+270 u^2}{6646(2\sigma)^3}.
\end{align}

\noindent To compare, in one of the simulations we obtained

\begin{align}
    b_2^{\textrm{exp}}=\frac{1737}{5000},\qquad r_2^{\textrm{exp}}=\frac{3263}{5000},
\end{align}

\noindent which differs from the $N\rightarrow\infty$ values, in the $0.01\%$ level. More distributions can be computed. However, the exact expressions, although polynomials in $u$ and $L$, have coefficients that are so big that this paper lacks space to include them here. A plot with $p_t(u)$ for $t=0,1,2,3,4$ is included bellow as well as $p_2(u)$ with the histogram obtained from the simulations:

\begin{figure}[H]
    \centering
    \begin{minipage}{0.47\textwidth}
        \centering
        \includegraphics[width=\textwidth]{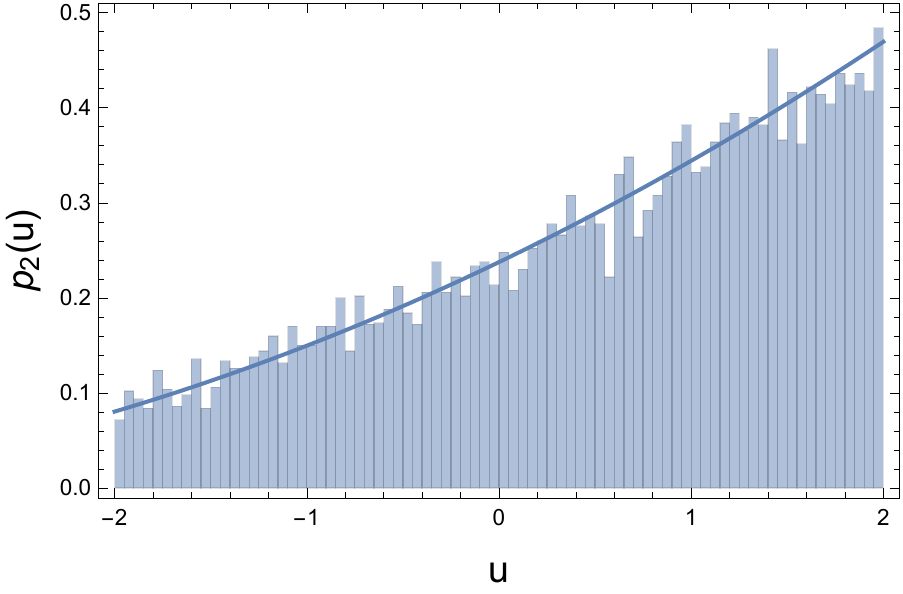}
        \caption{Distribution $p_2(u)$ Eq. (\ref{p2}) and histogram obtained in the simulation with $N=10.000$ and $\sigma=1$ at step $t=2$.}
    \end{minipage}\hfill
    \begin{minipage}{0.47\textwidth}
        \centering
        \includegraphics[width=\textwidth]{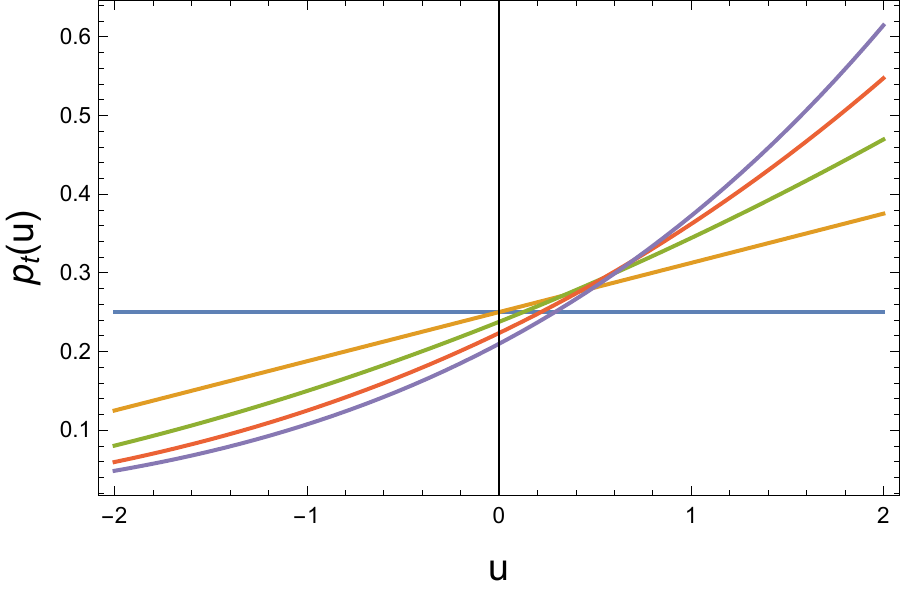}
        \caption{Distributions $p_t(u)$ with $t=0,1,2,3,4$ corresponding to the lines from the bottom to the the upper one at $u=2$.}
    \end{minipage}
\end{figure}

As mentioned in the previous section, we do not know how to rigorously prove that this system of equations for the evolution of $p_t(u)$ is indeed the $N\rightarrow\infty$ limit of the model we proposed in the previous section. However, the intuitive derivation, the self-consistent fact that the evolution preserves probability, and the good agreement with the simulations, strongly suggests that this is indeed the case. On the other hand, there are many details and subtleties, so it may be the case that there are higher-order effects that we are not considering.  As a final comment. It would be interesting to take the continuum limit in $t$ and find a differential equation that describes the evolution of the continuum probability distribution $p_{c}(u,t)$ and $p_{s}(u,t)$. Quantum mechanics deals with the unitary evolution of amplitudes. Perhaps, one could find a quantum mechanical analog of this system.

\subsection{Probability distribution for married couples}
\label{marriedcouplesdistribution}

In the model with marriage, married couples decouple from the system because they do not break. More precisely, they are not selected in the matching process as in Eq. (\ref{marriedcouples}). This sole fact, allows us to derive an expression for the utility distribution for married couples. This derivation will highlight the fact that this distribution is the same at all steps of the evolution of the system. As the system evolves, a larger fraction of agents will get married until all agents are married and then the system will not evolve anymore. Therefore, the asymptotic utility distribution for the system with marriage 

\begin{align}
    p_\infty(u)\equiv\lim_{t\rightarrow\infty}p_t(u),
\end{align}

\noindent is equal to the probability distribution of married couples at each step. In this section, we will find an analytical expression for this expression and compute it for the uniform model.

To start constructing this distribution first consider the agents that have an initial utility $u<\Lambda$. Eventually, they will meet an agent that will make their utility to be larger than $\Lambda$ and get marriage. Because the affinity with other agents is given by a distribution $p_0(u)$, the probability it will get married to an agent with affinity $u$ will end up being proportional to $p_0(u)$ but larger than $\Lambda$. Hence we will have

\begin{align}
    \frac{p_0(u)\Theta(u-\Lambda)}{\int_\Lambda^\infty p_0(s)ds}\int_{-\infty}^\Lambda p_0(x)dx
\end{align}

\noindent where the second-factor accounts to the volume of agents with $u<\Lambda$. For an agent with initial utility $u\geq\Lambda$, the same will hold, but now the final distribution needs to be larger than $u$ and not $\Lambda$, that is

\begin{align}
    \frac{p_0(u)\Theta(u-x)}{\int_x^\infty p_0(s)ds}.
\end{align}

\noindent Integrating for all values of $x$ larger than $\Lambda$ gives our ansatz for the final probability distribution:

\begin{align}
    p_\infty(u)=\int_\Lambda^\infty\frac{p_0(u)\Theta(u-x)}{\int_x^\infty p_0(s)ds}p_0(x)dx+\frac{p_0(u)\Theta(u-\Lambda)}{\int_\Lambda^\infty p_0(s)ds}\int_{-\infty}^\Lambda p_0(x)dx.
    \label{pinfinite}
\end{align}

\noindent Note that $p_\infty(u)$ is indeed normalized seems

\begin{align}
    \int_{-\infty}^\infty p_\infty(u)du=\int_\Lambda^\infty p_0(u)du+\int_{-\infty}^\Lambda p_0(u)du=1.
\end{align}

\noindent Recall that, $p_\infty(u)$ is the distribution for the married couples at each step of the evolution of the system.

For the Uniform model we have

\begin{align}
    p_\infty(u)=\ln \left(\frac{2\sigma-\Lambda }{2\sigma-u}\right)\Theta(u-\Lambda)p_0(u)+\frac{\Lambda+2\sigma}{2\sigma-\Lambda}\Theta(u-\Lambda)p_0(u),
    \label{pinfty}
\end{align}

\noindent where we used that

\begin{align}
    \int_\Lambda^\infty\frac{\Theta(u-x)}{\int_x^\infty p_0(s)ds}p_0(x)dx=\Theta(u-\Lambda)\int_{\Lambda}^u\frac{1}{2\sigma-x}dx=\ln\Big(\frac{2\sigma-\Lambda}{2\sigma-u}\Big)\Theta(u-\Lambda).
\end{align}

\noindent Visually, for $\sigma=1$, the distributions for different values of $\Lambda$ looks like

\begin{figure}[H]
    \centering
    \includegraphics[width=0.7\textwidth]{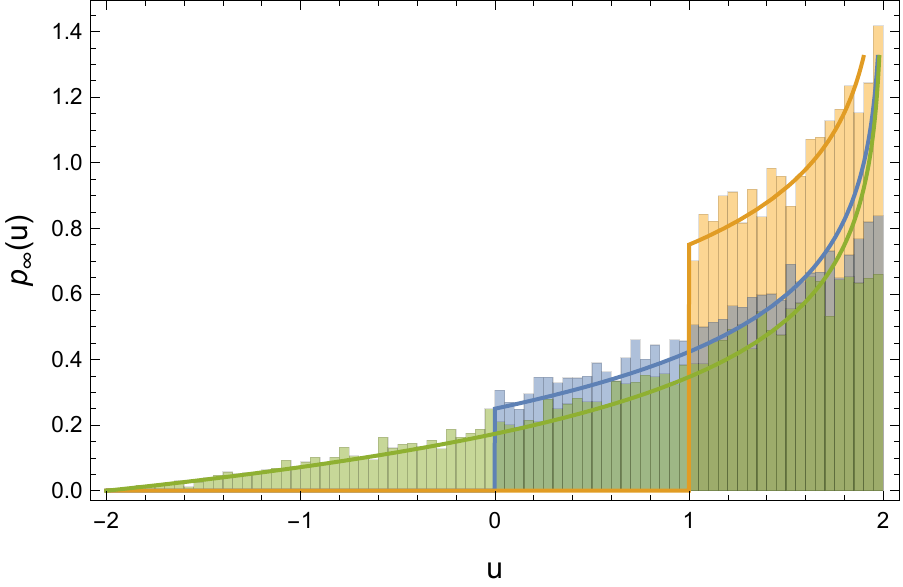}
    \caption{Asymptotic probability density function $p_\infty(u)$ for $\sigma=1$ and $\Lambda=1,0,-2$.}
\end{figure}

Note that the first term of $p_\infty(u)$ in Eq. (\ref{pinfty}) diverges as $u\rightarrow 2\sigma$, therefore, we need to regulate $p_\infty(u)$ to compute expectation values $E[U_\infty^n]$. This can be done by integrating from $-2\sigma$ to $2\sigma(1-\epsilon)$ instead of $2\sigma$, and then taking the limit $\epsilon\rightarrow0$. This regularization procedure gives

\begin{align}
    E[U_\infty^n]=\frac{ (2\sigma)^n }{2(n+1)}\big[H_{n+1}+\ln \big(1-\tfrac{\Lambda}{2\sigma} \big)+ B_{\frac{\Lambda }{2\sigma}}(n+2,0)\big]+\frac{\Lambda+2\sigma}{2\sigma-\Lambda}\frac{(2\sigma)^{n+1}-\Lambda^{n+1}}{2\sigma(n+1)}
\end{align}

\noindent where $H_n$ is the $n$-th harmonic number and $B_x(a,b)$ is the incomplete beta function. We compared these values with the sample moments obtained for the married couples at each step of the evolution of our system for different values of $\Lambda$ and    $\sigma$ and we obtained a good agreement. In particular, we have

\begin{align}
    E[U_\infty]=\frac{3 \Lambda ^2}{16 \sigma }+\frac{3 \Lambda }{4}+\frac{7 \sigma }{4},
    \label{asymptoticutilityuniform}
\end{align}

\noindent which gives $\sigma$ for $\Lambda=-2\sigma$. In the simulation without marriage with $N=10.000$ and $\sigma=1$ we got $u_{100}\simeq1.27$ which is a bit larger than the value $u_\infty=\sigma=1$. This result comes to reinforce the conclusions from the previous section (\ref{benefitsofmarriage}): don't marry the first person you like. In addition, the average utility Eq. (\ref{asymptoticutilityuniform}), as a function of $\Lambda$, attains its maximum value of $4\sigma$ for $\Lambda=2\sigma$. The value of $4\sigma$ for $\sigma=1$ is much larger than the values attained in our computational simulations in section (\ref{benefitsofmarriage}). Naively, this could suggest that the larger the value we establish to get married the better. But this is not true for steps $t\sim100$ as we showed with Fig. (\ref{utumtGn10000s1l1}) and Fig. (\ref{utumtUN10000s1l1}). This situation is a clear illustration of Keynes famous quote ``in the long run we are all dead" \cite{keynes1923tract}. Finally, it is nice to check that the expression for $E[U_\infty]$, is consistent with the dictionary explained in Eq. (\ref{dictionarymarried}). We indeed have

\begin{align}
    \sigma \times E[U_\infty]\Big\vert_{\sigma=1}=\frac{3\Lambda^2}{16}+\frac{3\sigma\Lambda}{4}+\frac{7\sigma}{4}=E[U_{\infty}]\Big\vert_{\Lambda=\sigma\Lambda}.
\end{align}





\section{Conclusions}

We proposed and investigated a model for mate searching and marriage in large societies based on a stochastic matching process and simple decision rules. By investigating this model, we disclosed some benefits of marriage as a search strategy. Most significantly, the average utility in the system with marriage can be higher than in the system without it. We summarized some of our results in what sounds like a piece of advice: don't marry the first person you like and don't search for the love of your life, but get married if you like your partner more than a sigma above average. We also mentioned that the average utility attained in our stochastic model is smaller than the one associated with a stable matching achieved using the Gale-Shapley algorithm. This can be taken as a formal argument in favor of a central planner (perhaps an app) with the information to coordinate the marriage market in order to set a stable matching. To roughly test the adequacy of our model to describe existent societies, we compared the evolution of the number of married couples between our model and real-world data and obtained an arguably good agreement. Lastly, we formulated the model in the limit of an infinite number of agents and found an analytical expression for the evolution of the system. 

This work left open many directions for investigation. For those with a mathematical bend, it would be relevant to prove that the evolution of our model in the large $N$ limit is indeed given by the evolution equations of section (\ref{analytical}). It would be interesting to explore and try to understand the $t\rightarrow\infty$ limit in the system without marriage for finite $N$ and also $N\rightarrow\infty$. For those interested in the adequacy of this model to describe reality, it would be important to do a more systematic comparison between the fraction of couples in our model and real societies, as we did in Fig. (\ref{mcsimulation}) and Fig. (\ref{mcrealdata}). One could try to model a mapping between steps of evolution and chronological time. By doing so, it would be easier to infer $\Lambda$ from the fraction of couples and one would be able to discover the value of $\Lambda$ for different societies. Combined with our study of the average utility as a function of $\Lambda$, summarized in Fig. (\ref{lcomp}), one could then advise the agents of a given society to be more or less selective to get married. For those interested in perfecting this model to describe reality, it would be relevant to investigate the fact that the affinity of a couple changes as time together passes. With a dynamical affinity, it would be possible to investigate new connections with real data, as the rate of divorce. 
As we have just shown, there are many open directions to advance the understanding of our stochastic matching model. Hopefully, this study will pave the way for future works.

\vspace{5mm}
\noindent
\textit{Acknowledgments:} This work was motivated by personal experiences with my girlfriend Lis, whom I thank for her love and companionship. As I am sure that our affinities are more than one sigma above the average, I hope the results presented here will convince her that we should get married, as they have convinced me. I would also like to thank Seth Koren for his valuable comments. 

\newpage
\bibliographystyle{unsrt}
\bibliography{bib}

\begin{thebibliography}{10}

\bibitem{10.2307/2312726}
D.~Gale and L.~S. Shapley.
\newblock College admissions and the stability of marriage.
\newblock {\em The American Mathematical Monthly}, 69(1):9--15, 1962.

\bibitem{knuthstable}
D.E. Knuth.
\newblock {\em Stable Marriage and Its Relation to Other Combinatorial
  Problems: An Introduction to the Mathematical Analysis of Algorithms}.
\newblock CRM proceedings \& lecture notes. American Mathematical Soc., 1997.

\bibitem{gusfield1989stable}
D.~Gusfield and R.W. Irving.
\newblock {\em The Stable Marriage Problem: Structure and Algorithms}.
\newblock Foundations of computing. MIT Press, 1989.

\bibitem{roth1992two}
A.E. Roth and M.A.O. Sotomayor.
\newblock {\em Two-Sided Matching: A Study in Game-Theoretic Modeling and
  Analysis}.
\newblock Econometric Society Monographs. Cambridge University Press, 1992.

\bibitem{ren2021matching}
Jing Ren, Feng Xia, Xiangtai Chen, Jiaying Liu, Mingliang Hou, Ahsan Shehzad,
  Nargiz Sultanova, and Xiangjie Kong.
\newblock Matching algorithms: Fundamentals, applications and challenges, 2021.

\bibitem{article}
Alvin Roth.
\newblock The economist as engineer: Game theory, experimentation, and
  computation as tools for design economics.
\newblock {\em Econometrica}, 70:1341--1378, 02 2002.

\bibitem{10.2307/1831130}
Gary~S. Becker.
\newblock A theory of marriage: Part i.
\newblock {\em Journal of Political Economy}, 81(4):813--846, 1973.

\bibitem{10.2307/1829987}
Gary~S. Becker.
\newblock A theory of marriage: Part ii.
\newblock {\em Journal of Political Economy}, 82(2):S11--S26, 1974.

\bibitem{10.2307/1837421}
Gary~S. Becker, Elisabeth~M. Landes, and Robert~T. Michael.
\newblock An economic analysis of marital instability.
\newblock {\em Journal of Political Economy}, 85(6):1141--1187, 1977.

\bibitem{doi:10.1146/annurev-economics-012320-121610}
Pierre-André Chiappori.
\newblock The theory and empirics of the marriage market.
\newblock {\em Annual Review of Economics}, 12(1):547--578, 2020.

\bibitem{10.2307/23045893}
Christopher~A. Pissarides.
\newblock Equilibrium in the labor market with search frictions.
\newblock {\em The American Economic Review}, 101(4):1092--1105, 2011.

\bibitem{10.1257/002205105775362014}
Richard Rogerson, Robert Shimer, and Randall Wright.
\newblock Search-theoretic models of the labor market: A survey.
\newblock {\em Journal of Economic Literature}, 43(4):959--988, December 2005.

\bibitem{doi:10.1146/annurev-economics-111809-125046}
Lones Smith.
\newblock Frictional matching models.
\newblock {\em Annual Review of Economics}, 3(1):319--338, 2011.

\bibitem{10.1257/jel.20150777}
Hector Chade, Jan Eeckhout, and Lones Smith.
\newblock Sorting through search and matching models in economics.
\newblock {\em Journal of Economic Literature}, 55(2):493--544, June 2017.

\bibitem{10.2307/2780247}
Dale~T. Mortensen.
\newblock Matching: Finding a partner for life or otherwise.
\newblock {\em American Journal of Sociology}, 94:S215--S240, 1988.

\bibitem{10.2307/2951279}
K.~Burdett and M.~Coles.
\newblock Marriage and class.
\newblock {\em The Quarterly Journal of Economics}, 112(1):141--168, 1997.

\bibitem{Burdett1999LongTermPF}
K.~Burdett and M.~Coles.
\newblock Long-term partnership formation: Marriage and employment.
\newblock {\em The Economic Journal}, 109:307--334, 1999.

\bibitem{10.2307/2999430}
Robert Shimer and Lones Smith.
\newblock Assortative matching and search.
\newblock {\em Econometrica}, 68(2):343--369, 2000.

\bibitem{10.2307/44955185}
Marion Goussé, Nicolas Jacquemet, and Jean-Marc Robin.
\newblock Marriage, labor supply, and home production.
\newblock {\em Econometrica}, 85(6):1873--1919, 2017.

\bibitem{RePEc:bpj:bejmac:v:advances.1:y:2001:i:1:n:5}
Robert Shimer and Lones Smith.
\newblock Matching, search, and heterogeneity.
\newblock {\em The B.E. Journal of Macroeconomics}, 1(1):1--18, 2001.

\bibitem{10.2307/2245639}
Thomas~S. Ferguson.
\newblock Who solved the secretary problem?
\newblock {\em Statistical Science}, 4(3):282--289, 1989.

\bibitem{notebook}
Davi B.~Costa.
\newblock Mathematica notebook.
\newblock Posted in the Wolfram community website:
  \url{https://community.wolfram.com/groups/-/m/t/2345375} on the 08/18/2021.
  Also available in the Wolfram cloud:
  \url{https://www.wolframcloud.com/obj/davicosta/Published/DatingMarriageModel.nb}.

\bibitem{mas1995microeconomic}
A.~Mas-Colell, P.E.A. Mas-Colell, W.M. D, M.D. Whinston, J.R. Green, C.~Hara,
  P.P.E.J.R. Green, I.~Segal, Oxford~University Press, and S.~Tadelis.
\newblock {\em Microeconomic Theory}.
\newblock Oxford student edition. Oxford University Press, 1995.

\bibitem{keynes1923tract}
J.M. Keynes.
\newblock {\em A Tract on Monetary Reform}.
\newblock Macmillan, 1923.

\bibitem{Multi-period}
Sangram~V. Kadam and Maciej~H. Kotowski.
\newblock Multi-period matching.
\newblock {\em Working Paper Series rwp15-030, Harvard University, John F.
  Kennedy School of Government}, 2015.

\bibitem{doval2021dynamically}
Laura Doval.
\newblock Dynamically stable matching, 2021.

\bibitem{Credibility}
Morimitsu Kurino.
\newblock Credibility, efficiency, and stability: A theory of dynamic matching
  markets.
\newblock {\em The Japanese Economic Review}, 71, 03 2008.

\bibitem{DAMIANO200534}
Ettore Damiano and Ricky Lam.
\newblock Stability in dynamic matching markets.
\newblock {\em Games and Economic Behavior}, 52(1):34--53, 2005.

\bibitem{doi:10.1080/0092623X.2016.1178675}
M.~L. Haupert, Amanda~N. Gesselman, Amy~C. Moors, Helen~E. Fisher, and
  Justin~R. Garcia.
\newblock Prevalence of experiences with consensual nonmonogamous
  relationships: Findings from two national samples of single americans.
\newblock {\em Journal of Sex \& Marital Therapy}, 43(5):424--440, 2017.
\newblock PMID: 27096488.

\bibitem{doi:10.2307/2061670}
Linda~J. Waite.
\newblock Does marriage matter?
\newblock {\em Demography}, 32:483--507, 1995.

\bibitem{owidmarriagesanddivorces}
Esteban Ortiz-Ospina and Max Roser.
\newblock Marriages and divorces.
\newblock {\em Our World in Data}, 2020.
\newblock https://ourworldindata.org/marriages-and-divorces.

\end{thebibliography}

\end{document}